# Multi-scale modeling of folic acid-functionalized TiO$_2$ nanoparticles for active targeting of tumor cells

Edoardo Donadoni[a], Paulo Siani[a], Giulia Frigerio[a], Cristiana Di Valentin[a,b,*]

[a]Dipartimento di Scienza dei Materiali, Università di Milano-Bicocca,
via R. Cozzi 55, 20125 Milano Italy
[b]BioNanoMedicine Center NANOMIB, University of Milano-Bicocca, Italy

**Abstract**

Strategies based on the active targeting of tumor cells are emerging as smart and efficient nanomedical procedures. Folic acid (FA) is a vitamin and a well-established tumor targeting agent because of its strong affinity for the folate receptor (FR), which is an overexpressed protein on the cell membranes of the tumor cells. FA can be successfully anchored to several nanocarriers, including inorganic nanoparticles (NPs) based on transition metal oxides. Among them, TiO$_2$ is extremely interesting because of its excellent photoabsorption and photocatalytic properties, which can be exploited in photodynamic therapy. However, it is not yet clear in which respects direct anchoring of FA to the NP or the use of spacers, based on polyethylene glycol (PEG) chains, are different and whether one approach is better than the other.

In this work, we combine Quantum Mechanics (QM) and Classical Molecular Dynamics (MD) to design and optimize the FA functionalization on bare and PEGylated TiO$_2$ models and to study the dynamical behavior of the resulting nanoconjugates in pure water environment and in physiological conditions. We observe that they are chemically stable, even under the effect of increasing temperature (up to 500 K). Using the results from long MD simulations (100 ns) and from free energy calculations, we determine how the density of FA molecules on the TiO$_2$ NP and the presence of PEG spacers impact on the actual exposure of the ligands, especially by affecting the extent of FA-FA intermolecular interactions, which are detrimental for the targeting ability of FA towards the folate receptor. This analysis provides a solid and rational basis for experimentalists to define the optimal FA density and the more appropriate mode of anchoring to the carrier, according to the final purpose of the nanoconjugate.

---

[*] Corresponding author: cristiana.divalentin@unimib.it

# 1. Introduction

Inorganic nanoparticles (NPs) based on transition metal (TM) oxides have been gaining much attention in recent years as theranostic agents for cancer therapy because of their versatile and unique physico-chemical properties and high biocompatibility.[1] Among these, titanium dioxide ($TiO_2$) NPs are very promising materials because they can act as photocatalysts for tissue oxidation processes or for the photodynamic therapy, under UV-visible light irradiation.[2]

Nanoparticles with proper dimensions (5-300 nm) can be internalized by tumor cells through the passive diffusion following the looser vascularization of the tumor tissues, also known as the enhanced permeability and retention (EPR) effect.[3] However, the non-specific accumulation of the NPs at the tumoral sites represents the major drawback of this kind of nanomedical approach, leading to undesired and potentially harmful side effects.[4] As a consequence, novel therapeutic strategies are based on the active targeting of the tumor cells, i.e. they use functionalized nanoparticles with ad-hoc biomolecules, which are capable of recognizing specific receptors that are overexpressed by the tumor cells.[5,6]

For instance folic acid (FA, also known as vitamin B9), whose chemical structure comprises a pterin portion, a *p*-aminobenzoate group and a glutamate residue (Fig. 1a), is recognized as a promising and efficient tumor targeting agent, because of a series of advantages, including its low molecular weight, water solubility, stability at different temperature and pH conditions, biocompatibility, good tissue penetration, excellent conjugation chemistry and, most of all, strong interaction with its receptor.[7] In particular, folic acid shows high affinity for the folate receptor (FR), a membrane protein that is expressed by neoplastic cells, such as those of epithelial, ovarian, breast, lung, kidney and brain tumors.[8] The folate cellular internalization is proven to happen via receptor-mediated endocytosis through endosome formation.[9]

A large number of nanosystems loaded with FA and drug molecules or contrast agents have been reported to be efficiently delivered to the tumor sites, and these include liposomes, polymeric micelles, dendrimers and inorganic NPs.[10–13] The latter, in particular, have superior efficacy for biofunctionalization, because of their high surface to volume ratio and greater surface modification potential with respect to organic NPs. Different TM oxide nanoparticles, including copper,[14] cobalt,[15] silicon,[16] and superparamagnetic iron[17] oxide NPs, have been successfully functionalized with folic acid, in order to combine the optical and/or catalytic properties of the inorganic nanoparticles with the targeting ability of FA, that is proven to be retained also after surface grafting.

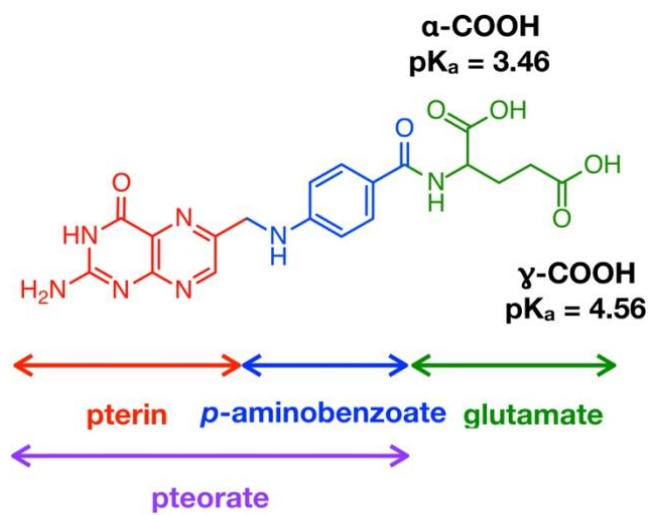

(a)

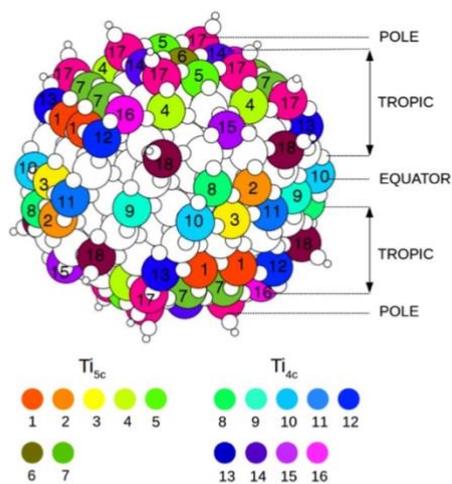

(b)

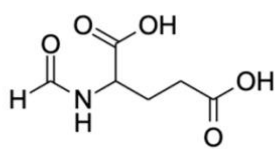

(c)

Fig. 1. (a) Chemical structure of folic acid (FA). (b) Graphical representation of the TiO$_2$ spherical nanoparticle model adopted in this work. The 4- and 5-fold Ti atoms are labelled with different colors and numbers and they are divided into sites that belong to the equator, to the tropics and to the poles of the nanoparticle. (c) Chemical structure of N-formylglutamic acid (FGA).

TiO$_2$/FA nanoconjugates have also shown an excellent ability to reach folate receptors in FR(+) cancer cells. For example, Ai et al.[18] report that TiO$_2$/FA nanoparticles have better capability of inducing reactive oxygen species (ROS) formation with higher cell apoptosis in osteosarcoma cancer cells than bare TiO$_2$ NPs, and that is related to the enhanced cellular uptake of these nanoconjugates mediated by FR in the presence of FA. Furthermore, Liang et al.[19] have combined the tumor targeting property of FA and the light harvesting capability of phthalocyanine (Pc) by synthetizing TiO$_2$/FA/Pc complexes for targeted photodynamic therapy (PDT). As an alternative, the light absorption frequency of TiO$_2$ can also be shifted from the UV to the visible range by its nitrogen doping.[20] Lai et al.[21] prepared FA-decorated TiO$_2$ NPs at different FA/TiO$_2$ weight ratios and found out that a value of 0.2 could yield nanoparticles having higher cytotoxicity under photoexcitation.

The strategy to use FA targeting ligand was successfully applied not only to PDT based on TiO$_2$ NPs but also on other inorganic photosensitizers, such as semiconducting black phosphorous nanosheets,[22] metal porphyrins[23] and metal phthalocyanines,[24,25] resulting in a more efficient cancer therapy.

It is also common practice in nanomedicine to decorate the nanoparticles with biocompatibility enhancing agents, such as polyethylene glycol (PEG) chains, that increase the blood circulation time of the nanosystems by shielding them from the activity of the reticuloendothelial system (RES), improving the NPs so-called "stealth properties".[26] Devanand et al.[27] succeeded in loading the antitumor drug paclitaxel on PEG-grafted and FA-functionalized TiO$_2$ NPs: the authors report a reduced paclitaxel loading on the nanoconjugates, because of the hindrance of PEG and FA, but an increased overall biocompatibility of the NPs and a sustained drug release phase. Moreover, their cytotoxicity assays demonstrated that the tumor cells' viability, induced by the nanoparticles, is concentration dependent. Naghibi et al.,[28] instead, combined the stealth properties provided by PEG, the targeting ability by FA and the tumor killing function by DOX on a TiO$_2$ NP. In particular, they found out that the optimal PEG/TiO$_2$ and FA/TiO$_2$ weight ratios are 2 and 0.5, respectively, to get the best targeting action against HFF2 human fetal foreskin fibroblasts and SKBR3 human breast cancer cells. At last, Shah et al.[29] synthetized cobalt- and/or nitrogen-doped bare or PEGylated TiO$_2$ nanoparticles, in the presence or without folic acid. They found out that the metal and non-metal doping boosted the photoactivation of the NPs in the Vis/NIR region, although the photokilling of the tumor cells was reduced with respect to the one induced by PEGylated undoped TiO$_2$ nanoparticles under UV/Vis radiation.

According to several experimental works, partly cited above, folate-targeted therapies have turned out to be remarkably successful approaches. However, to the best of our knowledge, only a few theoretical studies have dealt with the computational simulation of folic acid-tagged inorganic nanosystems as targeting nanosystems. In this work, for the first time, we have designed and built models of fully decorated folic acid-functionalized TiO$_2$ nanoparticles (~3000 atoms) by quantum chemical modeling, evaluating their thermal stability through simulated annealing calculations (section 3.1) and their dynamical behavior in water through molecular dynamics (MD) simulations by means of Classical Mechanics-based methods (section 3.2). Moreover, in section 3.3 we present the MD study of a PEGylated TiO$_2$ nanoparticle, where FA molecules are attached to the end group of some selected PEG chains, together with some free energy calculations, to determine the role played by a spacer on the FA exposure to the surrounding environment. At last, in section 3.4 we analyze the results from the MD simulations of some selected systems in a more realistic physiological environment.

## 2. Theoretical and computational methods

In this work, we adopted a combination of computational techniques that are based both on Quantum Mechanics (QM) and on Molecular Mechanics (MM).

### 2.1. Self Consistent Charge Density Functional Tight Binding (SCC-DFTB)

At the QM level, we used the Self Consistent Charge Density Functional Tight Binding (SCC-DFTB) method,[30] for full atomic relaxation and Born-Oppenheimer Molecular Dynamics (MD) simulations. DFTB is an approximated Density Functional Theory (DFT) approach, where the total energy of the system, $E_{tot}^{DFTB}$, is calculated as:

$$E_{tot}^{DFTB} = \sum_{i}^{occ} \langle \psi_i | \hat{H}_0 | \psi_i \rangle + \frac{1}{2} \sum_{\alpha,\beta}^{N} \gamma_{\alpha\beta} \Delta q_\alpha \Delta q_\beta + E_{rep}$$

(1)

where the first term, known as band structure term, is the sum of one-electron energies, deriving from the diagonalization of an approximated Hamiltonian matrix; the second term, called coulombic term, is calculated self-consistently using, in a tight binding approach, a minimal basis set for the expansion of the Kohn-Sham orbitals, $\psi_i$; the last, or repulsion energy term, contains the ion-ion and the exchange-correlation terms.

All the DFTB calculations were carried out using the DFTB+ open-source code.[31] We employed the MATORG parameterization set[32] for the pairwise interactions of the atoms of both the TiO$_2$ nanoparticle and the adsorbed FA molecules. The description of the hydrogen bonding was further improved with the inclusion of the empirical HBD correction[33] ($\zeta = 4$). Full geometry optimization calculations were performed through the Conjugate Gradient algorithm, where the forces were relaxed to less than $10^{-4}$ a.u. and the threshold for the convergence of the SCC procedure was set to $10^{-6}$ a.u.

Born-Oppenheimer molecular dynamics simulations were performed within the canonical ensemble (NVT). The Newton equations of motion were integrated with the Velocity Verlet algorithm,[34] using an Andersen thermostat,[35] a timestep of 0.5 fs and a SCC tolerance of $10^{-3}$ a.u. Initial velocities were assigned according to the Maxwell-Boltzmann distribution. Being the dimension of our systems challenging in the view of a quantum chemical description, our DFTB-MD simulations were limited to a total sampling time of 20 ps, which allowed an affordable computational cost.

The TiO$_2$ NP model was designed by our group in previous works[36,37] and consists of a spherical anatase nanoparticle, carved from the crystalline bulk anatase structure and fully relaxed, first at the DFTB level of theory with a simulated annealing procedure, followed then by a DFT-B3LYP optimization. The stoichiometry of the NP is $(TiO_2)_{223} \cdot 10H_2O$ and it is characterized by an equivalent diameter of 2.2 nm (700 atoms, Fig. 1b). In the simulations presented in this study we started from the DFTB-optimized NP to which we added as many N-formylglutamic acid (FGA, Fig. 1c) molecules as possible (52 or 48 depending on the COOH group used to anchor the surface) to reach the highest coverage admissible by the surface available Ti sites and the steric hindrance between attached molecules. FGA is a smaller molecule but analogous to FA. The FGA-functionalized models underwent some thermal treatment through MD simulations and were then fully relaxed at the DFTB level of theory. After that, we converted FGA molecules into FA and only performed partial DFTB geometry optimization (for further details see below).

The FGA or FA molecular adsorption energy on the spherical nanoparticle, $E_{ads}^{mol}$, is defined as:

$$E_{ads}^{mol} = \frac{E_{NP+n_{FGA\ or\ FA}} - (E_{NP} + n_{FGA\ or\ FA} E_{FGA\ or\ FA})}{n_{FGA\ or\ FA}}$$

(2)

where $E_{NP+n_{FGA\ or\ FA}}$, $E_{NP}$ and $E_{FGA\ or\ FA}$ are, respectively, the DFTB total energy of the whole system, of the bare NP and of a single FGA or FA molecule in the gas phase and $n_{FGA\ or\ FA}$ is the number of adsorbed FGA or FA molecules.

The TiO$_2$/PEG model was obtained by our group in previous studies,[38,39] where the nanoparticle was grafted with 50 PEG molecules and fully optimized through MD simulations and atomic relaxation at the DFTB level of theory with the MATORG parameterization set. On this model we added either 10 or 20 FA molecules, as randomly distributed as possible, by a covalent linkage through an ester bond involving the terminal C of PEG and the FA $\gamma$-carboxylic group. The FA-functionalized PEGylated TiO$_2$ systems were only partially relaxed at the DFTB level of theory.

**2.2. Molecular Mechanics-Molecular Dynamics (MM-MD)**

The Molecular Mechanics-Molecular Dynamics method is a classical approach, where the electrons are not described explicitly, and the total potential energy of the system is calculated as a function of a set of empirical parameters and potentials that constitute a force field (FF). In this work, all the classical MD simulations were carried out employing the LAMMPS (19 Aug 2019 version) open-source code.[40] The TiO$_2$ NP was described by an improved Matsui-Akaogi FF, reparametrized by Brandt and Lyubartsev,[41] while the CGenFF[42–44] was chosen for the adsorbed FA molecules and the TIP3P water molecules. We simulated several TiO$_2$/FA systems, with different FA coverages and binding modes to the NP, both in vacuum and in water. The TiO$_2$/FA topologies were generated by means of the Moltemplate[45] package for LAMMPS and the systems were immersed in a 100 X 100 X 100 Å$^3$ cubic water box, built up with the Packmol[46] software. During all the atomistic MD simulations, we fixed the position of the FA atoms directly bound to the NP at the DFTB geometry and we treated the nanoparticle as a rigid body, as previously done by some of us,[47] free to translate and rotate as a whole, fixing its internal degrees of freedom at the DFTB-optimized geometry through the RIGID package[48] in LAMMPS. The remaining atoms of the FA molecules and the solvent molecules, instead, were free to evolve in time, at 303 K (NVT ensemble), making use of a constant 2 fs timestep for the integration of the Newton's equations of motion, where the SHAKE algorithm[49] imposed holonomic constraints on all the covalent bonds involving hydrogen atoms. Periodic boundary conditions were used.

In the case of the TiO$_2$/PEG/FA systems, we fixed the position of the nanoparticle and of the binding hydroxyl group of the PEG chains at the DFTB geometry and we made use of a 135 X 135 X 135 Å$^3$ water box. Long-range electrostatic interactions were evaluated by the particle-particle particle-mesh (PPPM) solver, using a real-space cutoff of 12 Å. Short-range Lennard-Jones interactions were smoothly truncated with a 12 Å cutoff by means of a switching function applied beyond 10 Å. A hundred minimization steps ensured that no overlaps between the atoms occurred, then a 1 ns equilibration followed and finally the phase space was explored for a total production time of 100 ns.

**2.3. Simulation analysis**

Atomic radial distribution functions (rdf) were computed with the Radial Distribution Function plugin of VMD,[50] considering all the atoms at a given distance *r* from the reference and falling in a spherical crown with a thickness of 0.1 Å.

The hydrogen bonds analysis was carried out by means of the Hydrogen Bonds tool[50] provided by VMD. The criteria for the H-bonds classification were set as 3.5 Å for the maximum donor-acceptor distance and 30° for the maximum acceptor-donor-hydrogen angle.

Average interatomic distances and radius of gyration were computed exploiting the interdist tool, included in the open-source software LOOS 3.1.[51] In particular, for all the systems we analyzed:

- the distance between the N atom of the amine group of the FA pterin and the six-fold $Ti_{6c}$ atom in the center of the NP, $d\ N^{FA}$-$NP^{center}$
- the distance between the FA center of mass (com) and the closest Ti atom of the NP surface, $d\ com^{FA}$-$NP^{surface}$

For the $TiO_2$/PEG/FA systems, we also evaluated:

- the end-to-end distance of the PEG chains, $<h^2>^{½\ PEG}$, i.e. the mean distance between the first and the last heavy atom of each PEG chain (oxygen of the -OH head and carbon of the -$CH_3$ tail)
- the end-to-end distance of the PEG-FA chains, $<h^2>^{½\ PEG\text{-}FA}$, i.e. the average distance between the first and the last heavy atom of each PEG-FA chain (oxygen of the -OH head of PEG and nitrogen of the -$NH_2$ group of the FA pterin)
- the mean distance from the surface of the PEG chains, $MDFS^{PEG}$, and of the PEG-FA chains, $MDFS^{PEG\text{-}FA}$, i.e. the distance between the center of mass of each PEG residue and the closest Ti atom of the NP surface, for the PEG and the PEG-FA chains, respectively
- the radius of gyration of the PEG and the PEG-FA chains, $R_g^{PEG}$ and $R_g^{PEG\text{-}FA}$, i.e. the root-mean-square distance between each atom in the chain and the center of mass of the chain itself, normalized by the number of atoms:

$$R_g(r_i; r^{mean}) = \sqrt{\frac{1}{N}\sum_{i=1}^{N}|r_i - r^{mean}|^2}$$

(3)

The calculations of the non-bonding interaction energies (vdW and electrostatic) were performed through the USER-TALLY package, implemented in the LAMMPS code.[40]

The root-mean-square deviation (RMSD) of atomic positions was estimated using the VMD RMSD Trajectory Tool package.[50]

The diffusion coefficients, D, of the TiO$_2$/FA and TiO$_2$/PEG/FA nanosystems were estimated from the Einstein equation:

$$MSD = 2nDt$$

(4)

where MSD is the mean-square-deviation of atomic positions, n is the dimensionality of the diffusion and t is the simulation time. In particular, the diffusion coefficients were obtained by fitting of the MSD at intervals of the 100 ns MD trajectory where the dependency with t was linear, with the ordinary-least-squares (OLS) method.

The free energy profile to transfer one FA molecule, covalently bonded to a PEG chain (500 Da, 11 monomeric units) attached to the NP, from the NP surface (10 Å from the NP center) towards the farthest distance compatible with the PEG chain length (40 Å from the NP center) was determined by the Adaptive Biasing Force (ABF) method.[52] The mean force, $\langle F_\xi \rangle_\xi$, exerted along the transition coordinate, $\xi$, is related to the free energy first derivative according to:

$$\frac{dF}{d\xi} = \langle \frac{\delta U}{\delta \xi} - \frac{1}{\beta} \frac{\delta \ln |J|}{\delta \xi} \rangle_\xi = -\langle F_\xi \rangle_\xi$$

(5)

where |J| is the Jacobian determinant to transform the cartesian coordinates to generalized coordinates and $\langle F_\xi \rangle_\xi$ denotes the cumulative mean of the instantaneous force, $F_\xi$, in bins $\delta \xi$ wide. To obtain the free energy profiles, we followed a protocol similar to that of a previous work by some of us:[53] the reaction coordinate, defined as the z-component of the distance between the center of mass of the FA molecule and the center of the NP, was discretized into six consecutive windows, each 5 Å wide, at a z-distance from 10 to 40 Å with respect to NP center. To keep the FA molecule at the desired z-distance from the center of the NP, its center of mass was harmonically restrained in both the window's edges and the xy plane perpendicular to the transition coordinate with a force constant of 10 kcal/mol/Å$^2$. For every window, a total simulation time of 150 ns was reached, after a preliminary 10 ns equilibration phase. The nanoparticle and the binding hydroxyl group of the PEG molecules were fixed to the DFTB geometry all along the simulation, with the NP principal axis aligned to the cartesian axis. To obtain the LJ free energy profiles, the atomic partial charges of the considered FA

molecule were set to zero; in order to keep the whole system neutral, the charge of the two H atoms of the last -CH$_2$ unit of the PEG chain bonded to the same FA molecule were adjusted.

## 3. Results and discussion
### 3.1. Building up the TiO$_2$/FA models through quantum mechanical calculations
#### 3.1.1. Folic acid binding modes

First, we have investigated the possible adsorption modes of one folic acid molecule on a spherical model of a TiO$_2$ nanoparticle (equivalent diameter of 2.2 nm). One may expect that FA binds the surface Ti atoms through the carboxylic groups either in a monodentate or bidentate or chelated mode, through a dissociative or non-dissociative mechanism. Moreover, FA contains two different carboxylic groups, labeled as $\alpha$-COOH and $\gamma$-COOH groups in Fig. 1a, where the former is the most acidic, as demonstrated by the experimental pK$_a$ values,[54] and therefore it is the group that more easily would dissociate onto the NP surface. However, experimental works claim that the interaction of FA with the folate receptor is much stronger when the $\alpha$-COOH group is free and available for the ligand-receptor binding.[55] For this reason and for comparative purposes, in this work, we have considered both the carboxylic groups as potential anchoring groups on the nanoparticle. Therefore, we built up two different binding models: model $\alpha$ (TiO$_2$/n-FA-$\alpha$), where the FA molecules bind the nanoparticle through the $\alpha$-COOH and model $\gamma$ (TiO$_2$/n-FA-$\gamma$), where the anchoring bond involves the $\gamma$-COOH group. In addition, to speed up the QM calculations, we preliminary considered an analogue of the FA molecule, the N-formylglutamic acid (FGA, see Fig. 1c), which is shorter but, like FA, contains the glutamic acid portion used to anchor to the NP.

In Fig. 2, we compare four FGA binding modes involving the $\alpha$-COOH group and the equatorial four-fold Ti sites (Ti$^9$ and Ti$^{11}$ in Fig. 1b, which were proven to be the most reactive[56,57]) of the TiO$_2$ NP: undissociated monodentate (UM), dissociated monodentate (DM), bidentate (B) and chelated (C), using the FGA molecular adsorption energy (eqn (2)) as a criterion for the relative stability of these different adsorption modes. For all the dissociative mechanisms, the dissociated proton from FGA was transferred to the nearest two-fold oxygen atom (O$_{2c}$) of the NP.

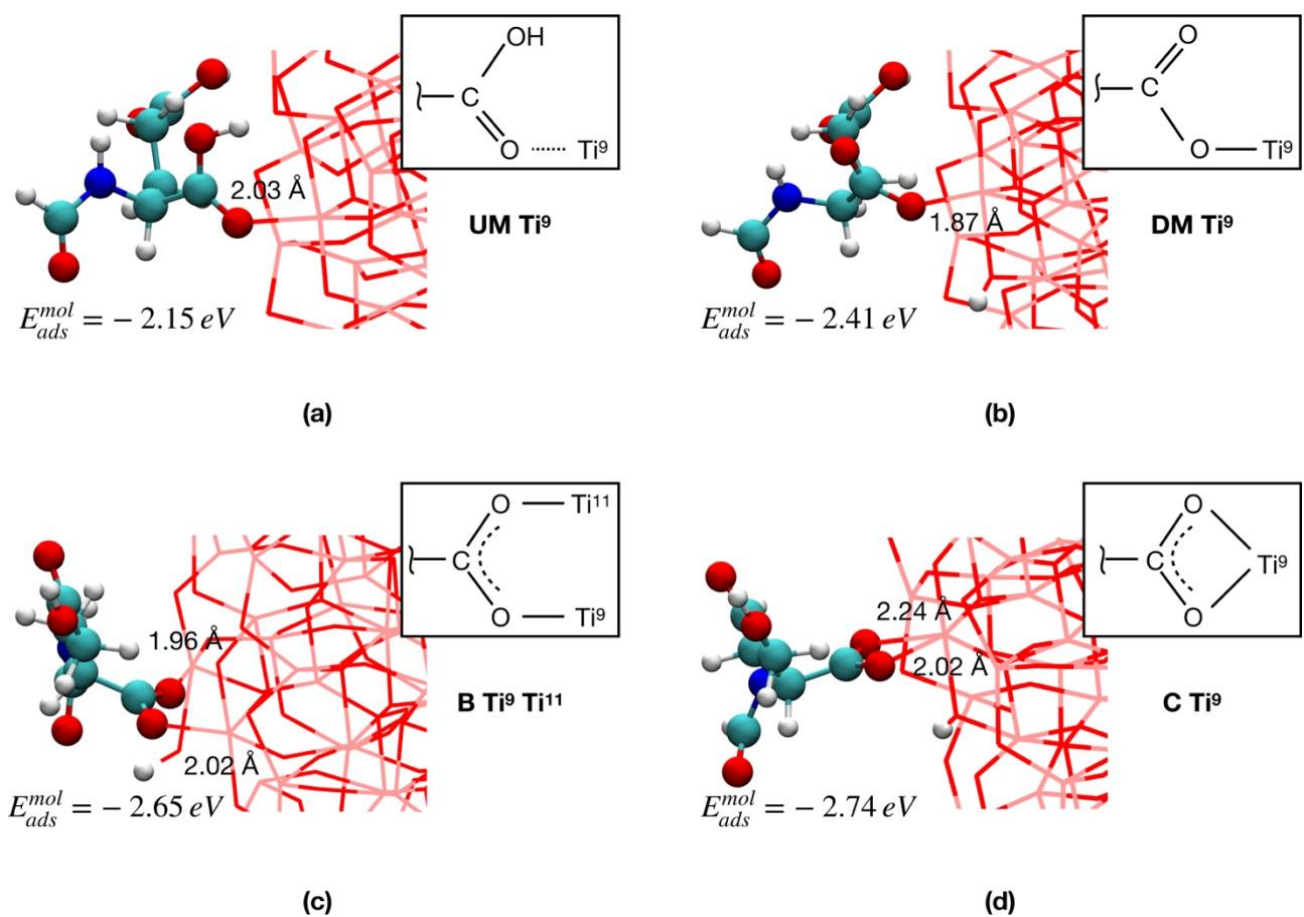

Fig. 2. Comparison of four FGA adsorption modes on the TiO$_2$ NP by the $\alpha$-COOH group: undissociated monodentate, UM (a), dissociated monodentate, DM (b), bidentate, B (c) and chelated, C (d). FGA molecular adsorption energies and NP-FGA Ti-O bond distances, computed at the DFTB level, are also reported. Titanium is shown in pink, oxygen in red, carbon in cyan, nitrogen in blue and hydrogen in white.

The results clearly show that the dissociative adsorption modes are always favored with respect to the non-dissociative ones: in particular, $E_{ads}^{mol}$ is 0.26 eV more negative for DM with respect to UM. Moreover, the B and the C coordination modes result, in the order, the most advantageous FGA binding modes with an $E_{ads}^{mol}$ that is, respectively, 0.24 and 0.33 eV more negative than that of DM.

On the basis of the results on FGA in this section, we conclude that, by extension, FA would preferentially have the anchoring carboxylic group dissociated on the TiO$_2$ surface, and that the ligand would likely form chelate coordinative bonds with Ti$_{4c}$ atoms and/or bidentate ones with Ti$_{4c}$ or Ti$_{5c}$ atoms, since these adsorption modes are energetically more stable. We rationalize these results as due to the greater energy stabilization by formation of two Ti-O bonds instead of one.

## 3.1.2. Increasing the coverage of the nanoconjugates

As a next step, we have proceeded with increasing the density of FGA molecules on the TiO$_2$ NP, starting with the equatorial Ti$_{4c}$ sites, based on our previous works where we demonstrated that these are the most reactive sites of the NP.[56,57] Hence, also considering the results of the previous section, we first saturated all the 12 equatorial Ti$_{4c}$ atoms with 12 FGA molecules in a chelated configuration. Fig. 3 shows the DFTB-optimized geometries of the TiO$_2$ NP decorated with an increasing n number of FGA molecules for the binding model $\gamma$ (TiO$_2$/n-FGA-$\gamma$) and Table 1 reports the average FGA molecular adsorption energy on the TiO$_2$ NP, at different FGA contents, for both binding models $\alpha$ (TiO$_2$/n-FGA-$\alpha$) and $\gamma$ (TiO$_2$/n-FGA-$\gamma$).

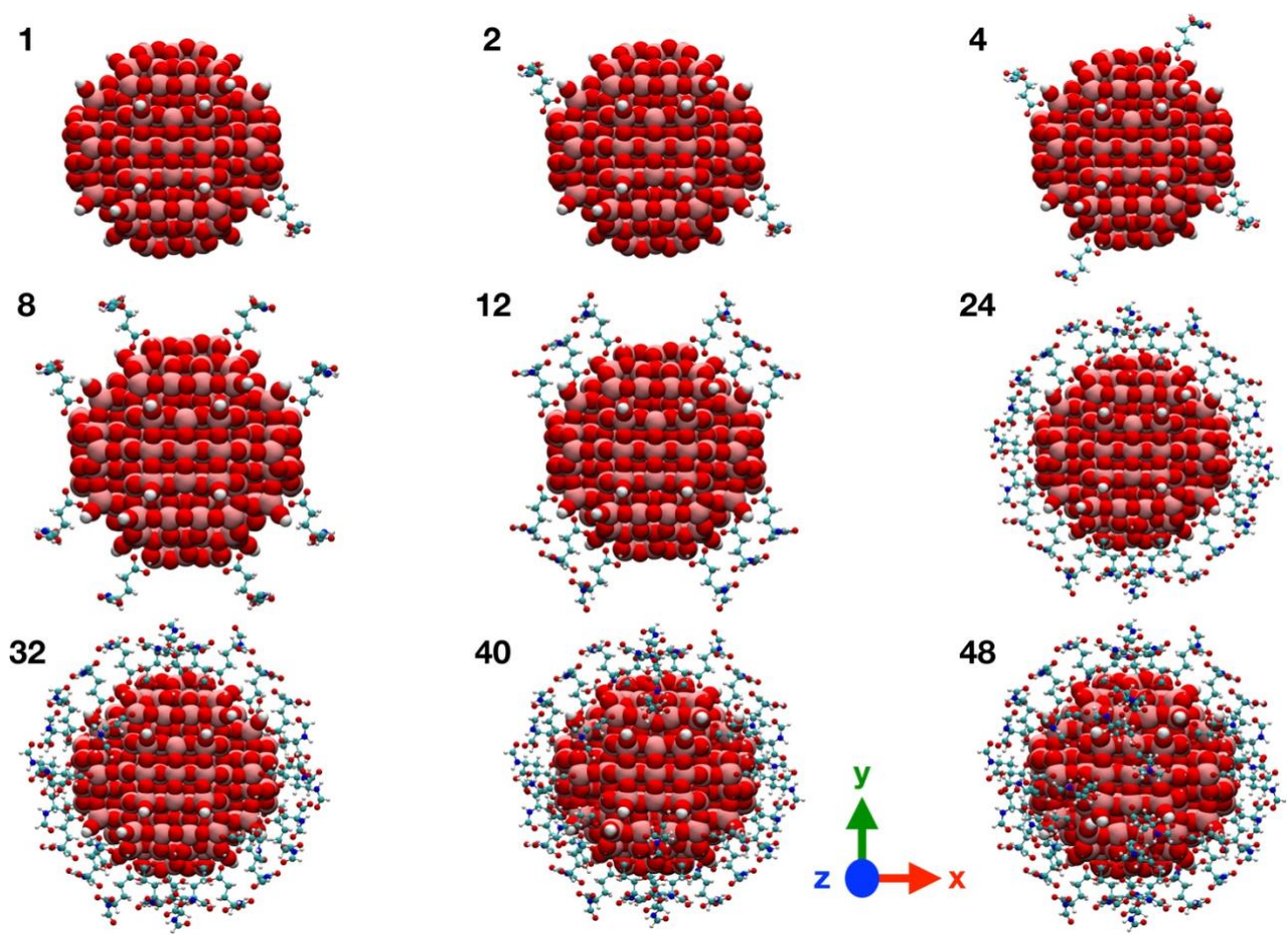

Fig. 3. Graphical representations of TiO$_2$/n-FGA-$\gamma$ system along the FGA covering process. Titanium is shown in pink, oxygen in red, carbon in cyan, nitrogen in blue and hydrogen in white.

Table 1. FGA molecular adsorption energies calculated at the DFTB level along the FGA covering process for binding models $\alpha$ (TiO$_2$/n-FGA-$\alpha$) and $\gamma$ (TiO$_2$/n-FGA-$\gamma$).

| FGA molecular adsorption energies | | | |
|---|---|---|---|
| TiO$_2$/n-FGA-$\alpha$ | | TiO$_2$/n-FGA-$\gamma$ | |
| $n_{FGA}$ | $E_{ads}^{mol}$ (eV) | $n_{FGA}$ | $E_{ads}^{mol}$ (eV) |
| 1 | -2.57 | 1 | -2.46 |
| 2 | -2.57 | 2 | -2.44 |
| 4 | -2.50 | 4 | -2.42 |
| 8 | -2.54 | 8 | -2.47 |
| 16 | -2.61 | 12 | -2.77 |
| 24 | -2.59 | 24 | -2.57 |
| 32 | -2.79 | 32 | -2.54 |
| 42 | -2.65 | 40 | -2.34 |
| 52 | -2.56 | 48 | -2.31 |

We observe from Table 1 that the addition of any further FGA molecule is accompanied by a decrease in the adsorption energy, although the effect is not large. This is true until the positions at the NP equator are filled in (12 FGA molecules, TiO$_2$/12-FGA). Then, we considered the tropics of the NP and added FGA molecules both in a chelated and a bidentate fashion: in particular, 16 chelated and 16 bidentate for binding model $\alpha$, whereas 8 chelated and 20 bidentate for binding model $\gamma$. These combinations were dictated by the general aim to get the highest coverage reducing the steric hindrance as much as possible. An increase in the coverage results in a larger adsorption energy, and that is due to the formation of intermolecular hydrogen bonds among the FGA molecules, which become more abundant with increasing adsorbed ligands. Finally, we saturated the Ti atoms at the poles with 8 FGA molecules, 4 chelated and 4 bidentate, in both models. At the end of the process, TiO$_2$/n-FGA systems are characterized by the following composition: binding model $\alpha$ has n = 52 FGA molecules (32 chelated and 20 bidentate), whereas binding model $\gamma$ has n = 48 FGA molecules (24 chelated and 24 bidentate).

Moreover, from the analysis of data in Table 1, one can notice that the FGA coordination through the $\alpha$-COOH (binding model $\alpha$) is energetically more advantageous, reflecting the greater acidity of the $\alpha$-COOH group, being closer to the electronegative N atom that exerts an electron-withdrawing polar effect. However, it is worth pointing out that the TiO$_2$/12-FGA-$\gamma$ system is characterized by a notably negative adsorption energy (-2.77 eV), that is the result of an extended network of intermolecular H-bonds among the amide groups of the FGA molecules (see Fig. S1), which would guide the arrangement of the ligands around the NP equator. Finally, comparing the adsorption energies at highest coverages, the FGA average adsorption energy for TiO$_2$/52-FGA-$\alpha$ results more negative (by 0.25 eV) with respect to that for TiO$_2$/48-FGA-$\gamma$.

Therefore, to conclude, although the coordination of any further FGA molecule is always energetically less convenient with respect to the adsorption of the first molecule, the intermolecular FGA-FGA H-bonds can make the average FGA adsorption energy more negative and, thus, contribute to the stabilization of the overall structure. Moreover, according to our results, the $\alpha$-COOH group of FGA is found to be a better anchoring group than $\gamma$-COOH.

### 3.1.3. Testing the thermal stability of the nanoconjugates

In this section, we evaluate the chemical stability of the TiO$_2$/52-FGA-$\alpha$ and TiO$_2$/48-FGA-$\gamma$ nanosystems by means of DFTB simulated annealing calculations. The simulated annealing is a computational technique, which allows to accelerate the exploration of the potential energy surface (PES) of a system, by heating it up to high temperatures and then slowly cooling it down, so that the system, during the process, is guided towards the global minimum of the PES. The protocol we followed for the simulated annealing calculations is the following: we heated up the TiO$_2$/52-FGA-$\alpha$ and TiO$_2$/48-FGA-$\gamma$ from 0 to 500 K for 1 ps, then we performed a 4 ps equilibration phase at constant temperature and, finally, we slowly cooled down the systems to 0 K in 15 ps. In Fig. S2, the temperature profiles that we adopted in the annealing process are shown. After this simulated annealing process, we took the last snapshot from each simulation and performed a further DFTB geometry optimization at 0 K with the Conjugate Gradient algorithm. Table 2 reports the average FGA molecular adsorption energies, at the beginning (OPT/DFTB (1)) and at the end (MD/DFTB) of the 20 ps simulation and also after the final DFTB geometry optimization (OPT/DFTB (2)). In addition, Fig. 4 presents a schematic and visual representation of the TiO$_2$/52-FGA-$\alpha$ system along the simulated annealing.

Table 2. FGA molecular adsorption energies calculated at the DFTB level for the TiO$_2$/52-FGA-$\alpha$ and TiO$_2$/48-FGA-$\gamma$ systems, at the beginning (OPT/DFTB (1)) and at the end (MD/DFTB) of the simulated annealing and after the final DFTB geometry optimization (OPT/DFTB (2)).

| FGA molecular adsorption energies | | |
|---|---|---|
| | $E_{ads}^{mol}$ (eV) | |
| | TiO$_2$/52-FGA-$\alpha$ | TiO$_2$/48-FGA-$\gamma$ |
| OPT/DFTB (1) | -2.56 | -2.31 |
| MD/DFTB | -2.57 | -2.30 |
| OPT/DFTB (2) | -2.66 | -2.37 |

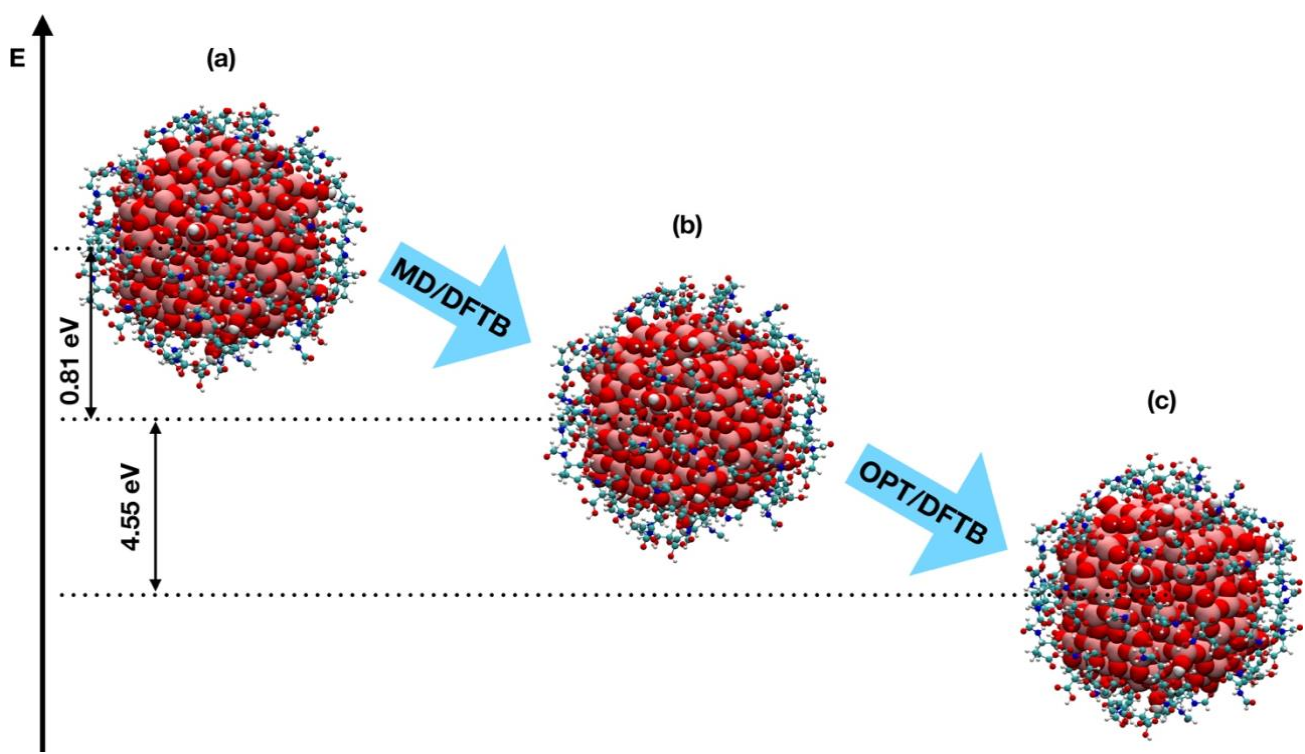

Fig. 4. Graphical and schematic representations of TiO$_2$/52-FGA-$\alpha$ system at the beginning (a) and at the end of the simulated annealing (b) and after the final DFTB geometry optimization (c). Titanium is shown in pink, oxygen in red, carbon in cyan, nitrogen in blue and hydrogen in white.

At a first sight, one would observe that the overall arrangement of the ligands around the nanoparticle is not significantly perturbed during the simulation. Nevertheless, as an indicator of the chemical stability of the nanoconjugates, we computed the average distances between the carbon atom of the FGA carboxyl group ($\alpha$-COOH for TiO$_2$/52-FGA-$\alpha$ system and $\gamma$−COOH for TiO$_2$/48-FGA-$\gamma$) and the corresponding coordinated Ti atom of the NP, as a function of the simulation time: the trends are shown in Fig. S3. The plots indicate that the mean C-Ti distances decrease during the heating step, they fluctuate more in the following equilibration phase and, finally, they increase again along the cooling steps. These distance values are compatible with the formation of a bond between the carboxylic group of FGA and a Ti atom of the nanoparticle, therefore we can conclude that the O-Ti bonds between FGA and the NP Ti atoms are preserved all over the simulated annealing simulation.

The observed chemical stability agrees with thermogravimetric analysis (TGA) by Devanand et al.[27] on PEGylated FA-functionalized TiO$_2$ NPs and by Mallakpour et al.[58] on FA-decorated TiO$_2$ NPs dispersed into a poly(vinyl alcohol) (PVA) polymeric matrix, where FA decomposition occurs in a temperature range of 200-800 °C.

Another quantitative parameter that we could extract from the simulations is the number of established hydrogen bonds: in Fig. S4 we report the time evolution of the H-bonds between FGA-NP fragments and FGA-FGA fragments. Moreover, the advantage of a Born-Oppenheimer MD relies on the fact that during the simulation it is possible to observe bonds breaking and formation and indeed, during the annealing, we could register a proton transfer in TiO$_2$/52-FGA-$\alpha$ system, from the free carboxylic group of one FGA molecule to a near O$_{2c}$ atom of the NP, with the formation of one NP-FGA Ti-O bond in a monodentate configuration (Fig. S5).

From the results in this section, we conclude that the O-Ti bonds between FGA and the TiO$_2$ NP are stable, even at high temperatures, such as those simulated in the annealing process, and that some proton transfers from FGA to TiO$_2$ may spontaneously occur.

**3.1.4. TiO$_2$/FA high coverage models**

Once the chemical stability of the TiO$_2$/52-FGA-$\alpha$ and TiO$_2$/48-FGA-$\gamma$ nanoconjugates has been established, we moved on by converting the FGA ligands into FA ones by adding the remaining portion of all the molecules, manually and simultaneously. At the end, thus, TiO$_2$/n-FA systems have the following composition: binding model $\alpha$ has n = 52 FA molecules (32 chelated and 20 bidentate for a total of 3351 atoms and with a FA/TiO$_2$ weight ratio of 1.28) and binding model $\gamma$ has n = 48 FA molecules (24 chelated and 24 bidentate for a total of 3147 atoms and with a FA/TiO$_2$ weight ratio of 1.18). After building up the models, we performed some DFTB geometry optimization steps with the Conjugate Gradient algorithm to prevent possible overlaps among the manually added FA molecules. Even if we could not proceed to reach full convergence because of the large number of atoms of the systems, these few additional optimization steps were sufficient to adjust the conformation and the spatial arrangement of the FA pteorates (see Fig. 1a), since the portion of the ligands anchoring the NP had been already optimized. In Fig. 5, the graphical representations of prepared TiO$_2$/52-FA-$\alpha$ and TiO$_2$/48-FA-$\gamma$ systems are shown.

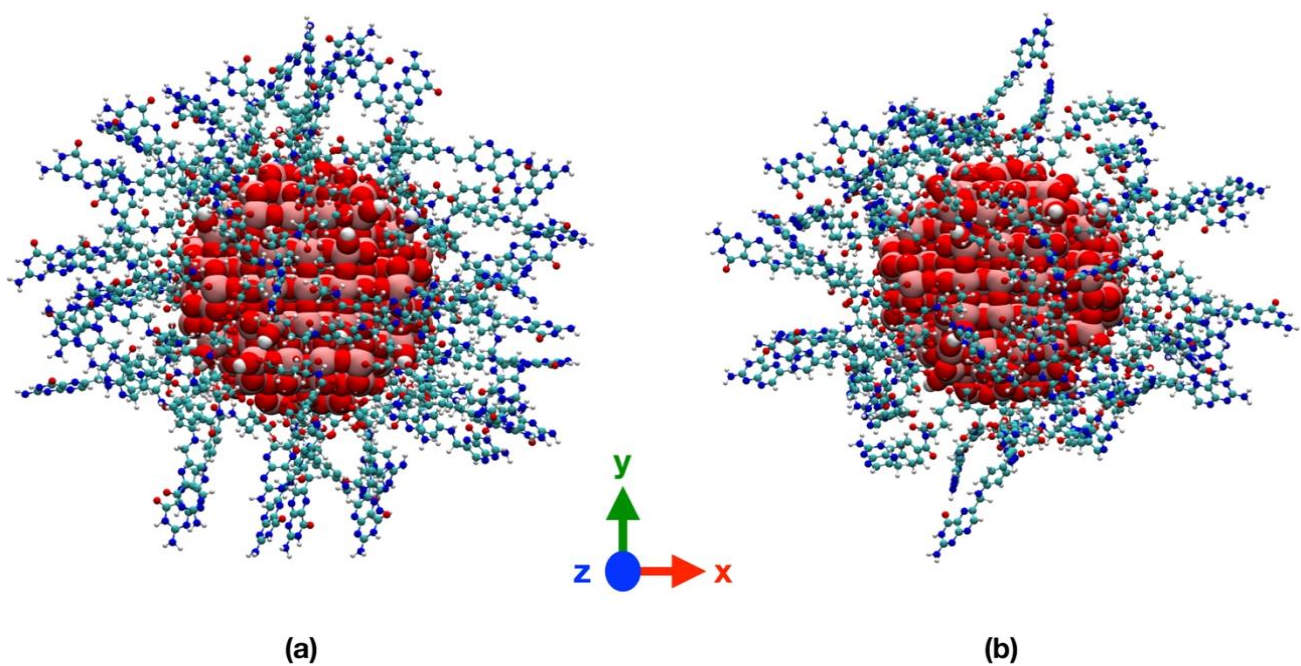

Fig. 5. Graphical representations of TiO$_2$/52-FA-$\alpha$ (a) and TiO$_2$/48-FA-$\gamma$ (b) systems after some DFTB geometry optimization steps on the by-hand-built structures. Titanium is shown in pink, oxygen in red, carbon in cyan, nitrogen in blue and hydrogen in white.

### 3.2. Classical MD simulations of TiO$_2$/FA systems
### 3.2.1. MD simulations of a TiO$_2$ NP functionalized with one FA molecule

The classical MD simulations were first performed on two simple systems, i.e. the TiO$_2$ NP functionalized with one single FA molecule, bound to an equatorial Ti$_{4c}$ site (Ti$^{10}$) in the chelated mode, either by the $\alpha$-COOH or the $\gamma$-COOH group. These two structures were obtained by substituting the FGA molecule in the corresponding DFTB-optimized TiO$_2$/1-FGA structures. In Fig. 6 we report the last snapshots of the 100 ns production MD simulations of the two TiO$_2$/1-FA systems, at 303 K, in vacuum and in water.

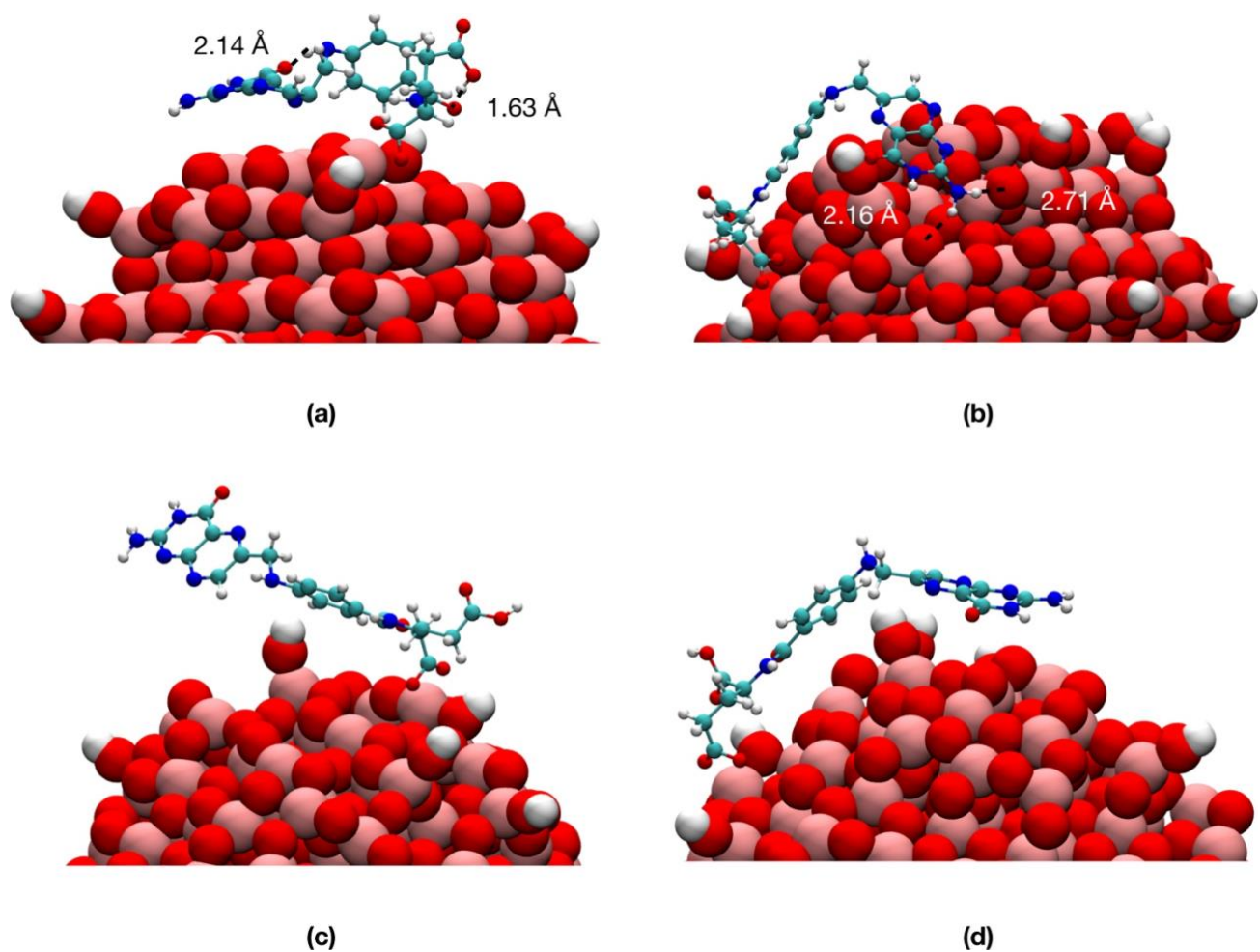

Fig. 6. Last snapshots from the 100 ns production simulations of TiO$_2$/1-FA-$\alpha$ in vacuum (a) and in water (c) and TiO$_2$/1-FA-$\gamma$ in vacuum (b) and in water (d). Titanium is shown in pink, oxygen in red, carbon in cyan, nitrogen in blue and hydrogen in white. The water molecules are not shown for clarity.

We observe that, in vacuum, FA stays closer to the NP (Fig. 6a,b) with the pterin bent to the TiO$_2$ surface, forming H-bonds with the NP in Fig. 6b and intramolecular H-bonds in Fig. 6a. On the contrary, in water, the ligand is more involved in interactions with the solvent (Fig. 6c,d).

In Table S1 we present the average values of some structural parameters we calculated along the 100 ns simulations, including atomic distances, hydrogen bonds and non-bonding interaction energies. In water, both the center of mass of FA and the terminal amine group of the pterin lie at greater distances from the NP surface and center, respectively, with respect to vacuum. The NP-FA vdW interactions are comparable in vacuum and in water, but that is not true for the electrostatic ones. Indeed, we observe a substantial decrease in the magnitude of the NP-FA electrostatic interactions in water, as the solvent competes with the ligand for the interactions with the NP: this is demonstrated by the magnitude of the NP-water and FA-water electrostatic interactions. On the contrary, in vacuum, the NP-FA electrostatic interactions are stronger and for the TiO$_2$/1-FA-$\gamma$ system they become even more relevant than the vdW ones.

### 3.2.2. MD simulations of a TiO$_2$ NP functionalized with 8 FA molecules

In this section, we analyze the TiO$_2$ nanoparticle that was functionalized with 8 folic acid molecules, corresponding to a FA/TiO$_2$ weight ratio of about 0.2. We focus the attention on this coverage density because it was proposed by Lai et al. to produce nanoconjugates with higher cytotoxicity under photoexcitation.[21]

We considered two systems where all the equatorial four-fold Ti[10] and Ti[11] sites are saturated by 8 FA molecules, which chelate either by the $\alpha$-COOH or the $\gamma$-COOH, respectively. These two structures were obtained from the corresponding TiO$_2$/8-FGA-$\alpha$ and TiO$_2$/8-FGA-$\gamma$ DFTB-optimized geometries by the substitution of the eight FGA molecules with FA ones. In Fig. S6 the last snapshots of the 100 ns production MD simulations, in vacuum and in water at 303 K, are shown. Along the whole simulation and for both models, the ligands stick to the surface and they self-interact to a greater extent in vacuum than in water, where they can establish favorable interactions with the solvent molecules.

To quantify all these interactions, in Table S2 we list the average atomic distances, hydrogen bonds and non-bonding interaction energies along the 100 ns MD simulations. Once again, in vacuum the FA molecules are found to keep closer to the NP surface, and that is supported by the higher number of NP-FA H-bonds. Also, the interaction among ligands is more pronounced in vacuum, as indicated by the magnitude of the FA-FA H-bonds. The FA-FA interactions are mostly of an electrostatic nature. On the contrary, in water, the FA-FA vdW interactions become predominant in the TiO$_2$/8-FA-$\gamma$ system and they are the highest (i.e. less favorable) in the TiO$_2$/8-FA-$\alpha$ system, where they are comparable to the electrostatic interactions. Regarding the NP-FA interactions, in all the cases they are predominantly of hydrophobic type. In water we also observe a decrease in the magnitude of the NP-FA vdW interactions and a more significant drop in the magnitude of the electrostatic interactions, compensated by stabilizing ligand-solvent interactions.

### 3.2.3. MD simulations of high coverage TiO$_2$/FA systems

At last, we discuss the MD simulations of the high coverage TiO$_2$/52-FA-$\alpha$ and TiO$_2$/48-FA-$\gamma$ systems, which have been partially optimized at the DFTB level (see Fig. 5). Fig. 7 shows the last snapshots from the 100 ns MD simulations of the two systems, in vacuum and in aqueous environment at 303 K.

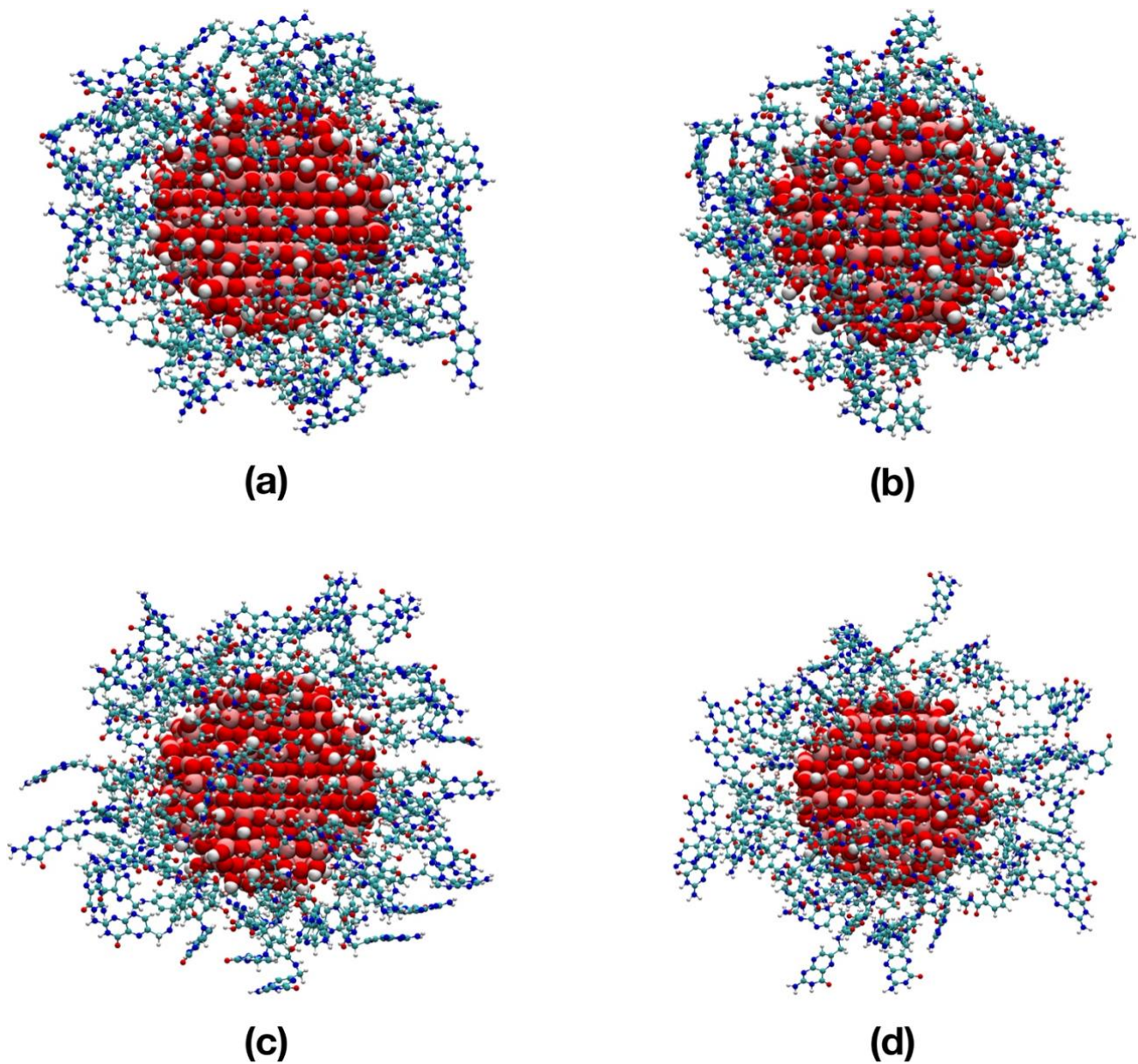

Fig. 7. Last snapshots from the 100 ns production simulations of TiO$_2$/52-FA-$\alpha$ in vacuum (a) and in water (c) and TiO$_2$/48-FA-$\gamma$ in vacuum (b) and in water (d). Titanium is shown in pink, oxygen in red, carbon in cyan, nitrogen in blue and hydrogen in white. The water molecules are not shown for clarity.

Despite the high ligand density, when in vacuum again, the FA molecules bend towards the NP surface in order to maximize the interactions, whereas, in water we observe a competition between the FA-NP and the FA-water interactions.

To quantitatively evaluate this consideration that comes from a first visual inspection, in Table S3 the average atomic distances, hydrogen bonds and non-bonding interaction energies along the 100 ns MD simulations are reported. We notice that the distance of the amine group of the pterin from the center of the nanoparticle only slightly increases when switching from vacuum to an aqueous environment. In line, we do not register an effective variation in the NP-FA H-bonds. What makes the difference, instead, is the extent of the FA-FA H-bonds, which decrease of about 75% from vacuum to water (the greater numbers for the TiO$_2$/52-FA-$\alpha$ system are due to the intrinsic higher number of adsorbed FA molecules). Indeed, the FA-FA electrostatic energy becomes less negative in water, as well as the NP-FA electrostatic energy. Finally, with respect to the TiO$_2$/8-FA systems, the FA-water interactions (per FA molecule) decrease from 115-113 kcal/mol ($\alpha$ and $\gamma$ in Table S2, respectively) to 95-99 kcal/mol ($\alpha$ and $\gamma$ in Table S3, respectively). These results clearly indicate a more effective exposure of FA molecules at low coverage with FA/TiO$_2$ weight ratio of 0.2, in line with the experimental findings by Lai et al.[21]

The increase in the total FA-water interaction energy (van der Waals and electrostatic, Table S1-S3) at increasing FA density on the TiO$_2$ NP suggests that the water solubility, and thus dispersibility, of the nanosystems improves as the number of adsorbed ligand molecules increases. Indeed, according to Xie et al.,[20] the surface coating of TiO$_2$ NPs with folic acid determines an increase in the magnitude of their zeta potential, up to a value that is close to the threshold of 30 mV, above which NPs are considered to be stably dispersed in the solvent.[59]

To compare all the TiO$_2$/FA systems considered so far, in Fig. 8 we report the rdf of the FA center of mass with respect to the six-fold coordinated Ti atom at the center of the NP, for each model, averaged over the 100 ns of the MD simulations.

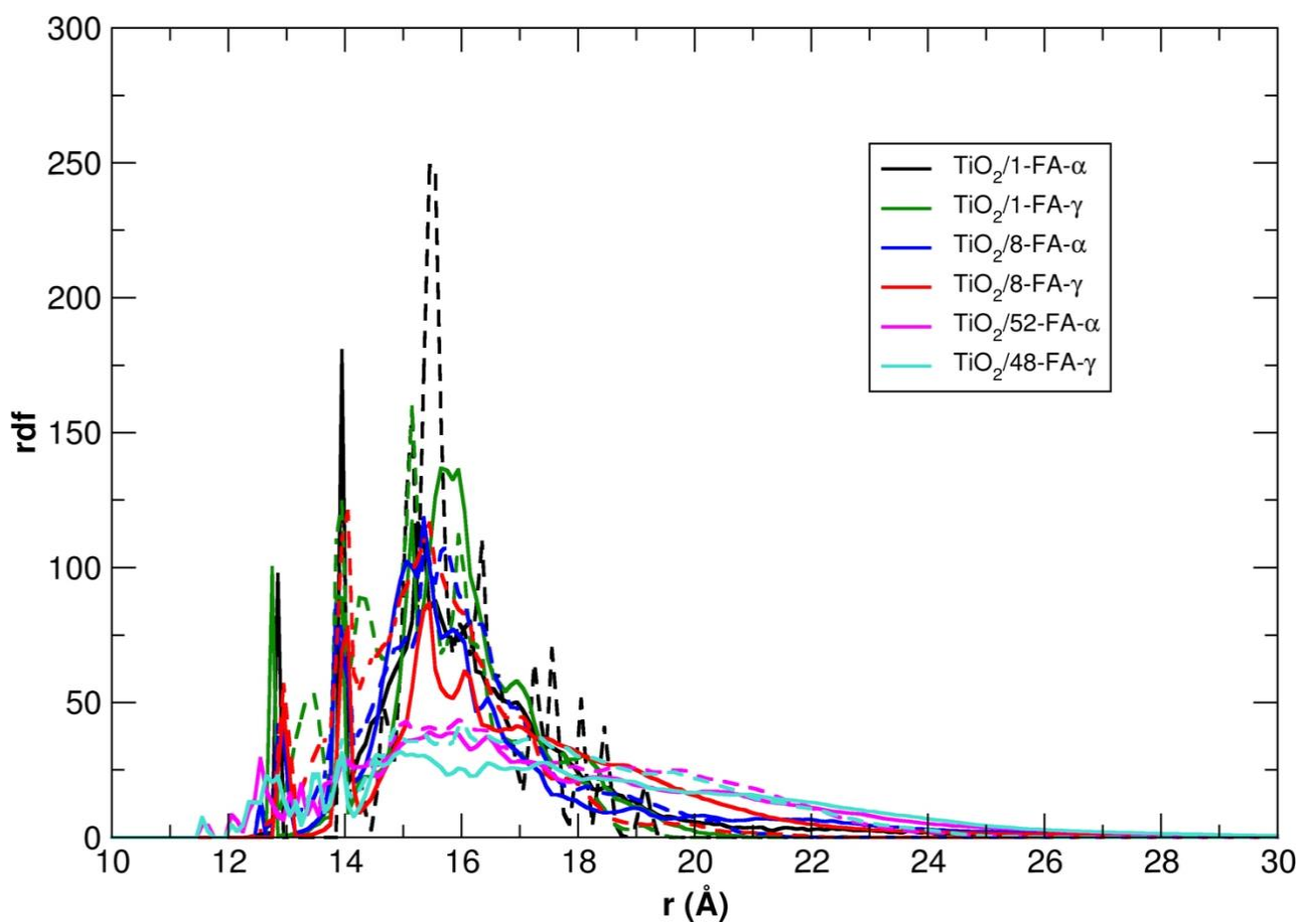

Fig. 8. Radial distribution function (rdf) of the FA center of mass with respect to the six-fold coordinated Ti atom at the center of the NP, for all the TiO$_2$/FA models, averaged over the 100 ns MD simulations in vacuum (dashed lines) and in water (solid lines).

The shape of the rdf curves becomes broader with the increase in the density of adsorbed folic acid molecules, whereas the maximum intensity, located at about 15-16 Å from the NP center, decreases. In addition, the radial distribution of the ligands around the nanoparticle is always broader in water than in vacuum, because the FA molecules have more potential stable conformations due to the interaction with the solvent molecules.

Finally, we study the dynamical properties of the nanoconjugates in an aqueous environment. In Fig. S7, the root-mean-square deviation of atomic positions (RMSD) plots of the NP+FA systems along the 100 ns MD simulations are shown with respect to the 0 ns-reference atomic positions. The slope of the RMSD vs time curves is related to the diffusion coefficient of the system. In particular, the slope of the curve relative to the TiO$_2$ systems grafted with a single folic acid molecule (TiO$_2$/1-FA-$\alpha$) is the highest, which means that this nanoconjugate diffuses through the aqueous medium at the highest velocity. Then, for the other systems, we observe lower slopes of the curves mostly due to the increase in the number of the adsorbed ligands, as the total mass of the NP+FA moiety increases.

The diffusion of the nanoconjugates is a crucial aspect to be studied, as the functionalized nanoparticles, after injection, must reach their cellular target, thus the rate of the transport will influence the efficiency (and therefore, the efficacy) of the nanomedical process. To this purpose, in Table S4 we also report the estimated diffusion coefficients of all the TiO$_2$/n-FA-$\alpha$ systems under investigation in water, computed by the OLS method at MD trajectory intervals showing a linear dependence between the MSD of atomic positions and time (see section 2.3 for more details). We observe that the values for all systems are of the same order of magnitude (i.e. 10$^{-10}$ m$^2$/s). Moreover, the diffusion coefficients decrease with increasing FA density on the NP.

Overall, the computed diffusion coefficients qualitatively agree with the experimental findings of Domingos et al.,[60] who determined the diffusion coefficients of bare and organic acid-functionalized TiO$_2$ nanoparticles (with a diameter of 5 nm) being in the order of magnitude of 10$^{-11}$ m$^2$s$^{-1}$.

### 3.3. Classical MD simulations of TiO$_2$/PEG systems functionalized with FA molecules
### 3.3.1. Preparation of the models

In this section, we study the dynamics of a TiO$_2$ nanoparticle grafted with PEG chains which are further functionalized with FA molecules. The PEG chain considered in this work is a methoxy-PEG, H$_3$C−[OCH$_2$CH$_2$]$_n$−OH, with n = 11, for a total molecular weight of 516 g/mol. One FA is covalently bonded to one of the PEG ends through an ester bond involving the FA $\gamma$-COOH. In Fig. S8, the DFTB-minimum energy geometry of a PEG-FA chain is shown: the optimized O$^{\gamma\text{-COOH(FA)}}$-$C^{-CH_3(PEG)}$ bond distance is 1.45 Å. Then, on the DFTB-optimized TiO$_2$/PEG geometry,[38,39] we substituted either 10 or 20 PEG chains with as many PEG-FA chains, randomly and homogeneously distributed, to obtain the TiO$_2$/PEG/10-FA and the TiO$_2$/PEG/20-FA systems, respectively. We used these compositions, since Naghibi et al.[28] reported that the performance of folic acid on the selective targeting of the cancer cells was maximized at a FA/TiO$_2$ weight ratio of 0.5, which is close to that obtained with the TiO$_2$/PEG/20-FA model. Moreover, we also built up the TiO$_2$/PEG/10-FA system, with a FA/TiO$_2$ weight ratio of 0.25, that is close to the optimal value proposed in another work by Lai et al.[21] After a thousand geometry optimization steps, performed at the DFTB level, we obtain the two models shown in Fig. S9.

**3.3.2. MD simulations of TiO$_2$/PEG/10-FA and TiO$_2$/PEG/20-FA systems**

In Fig. 9 we present the last snapshots at the end of the 100 ns MD simulations of the TiO$_2$/PEG/10-FA and TiO$_2$/PEG/20-FA systems, in vacuum and in water at 303 K. In vacuum, the PEG molecules interact among themselves and wrap around the nanoparticle, creating a barrier that shields the terminal FA ligands from the NP surface. On the contrary, in water, the PEG chains are more stretched, with their ends pointing to the aqueous phase and they interact less with the other PEG molecules, so that folic acid is more likely to be closer to the NP surface and have stronger interactions with the carrier.

To support our findings, in Table S5 we present an analogue structural analysis that we performed for the TiO$_2$/n-FA systems, evaluating several structural indicators averaged over the 100 ns MD simulations of the TiO$_2$/PEG/10-FA and TiO$_2$/PEG/20-FA systems. We confirm that in vacuum the FA molecules are closer to the TiO$_2$ surface than in water on the basis of the shorter distances between the amine group of the FA pterin and the Ti atom at the center of the NP, as well as the shorter distances of the FA center of mass from the closest Ti atom on the NP surface and the MDFS of the PEG-FA chains or, alternatively, on the basis of the more stabilizing NP-FA vdW and electrostatic interaction energies. Moreover, there is not a big difference in the values of the distance indicators that we obtained for the TiO$_2$/PEG/10-FA and in the TiO$_2$/PEG/20-FA systems. The values of the PEG and PEG-FA radius of gyration reflect the tendency of the PEG polymer chains to surround the NP, i.e. reduced R$_g$ numbers, in line with the end-to-end distances. As a consequence, the NP-PEG and especially the PEG-PEG interaction energies are more negative in vacuum than in water. Finally, it is interesting to point out that the FA-FA interactions play a significant role in the TiO$_2$/PEG/20-FA systems, affecting the interaction energies more than the H-bonds, where the FA-FA vdW energy is respectively 10 and 5 times greater in magnitude both in vacuum and in water compared to the TiO$_2$/PEG/10-FA systems, whereas the FA-FA electrostatic energy is even 400 times more negative in vacuum but just 6 times more negative in water than for the TiO$_2$/PEG/10-FA systems. A similar behavior was reported by us in another work[61] for another targeting moiety, i.e. cyclic RGD peptide (cRGD), which we investigated at comparable ligand densities to those considered here for FA and conjugated to the same PEGylated TiO$_2$ nanoparticle.

The FA-FA interactions could play a critical role in the recognition process of FA by the folate receptor on the cell membrane. The FA ligand must well fit into the protein's binding pocket, which could be hampered or prevented by excessive FA-FA interactions.

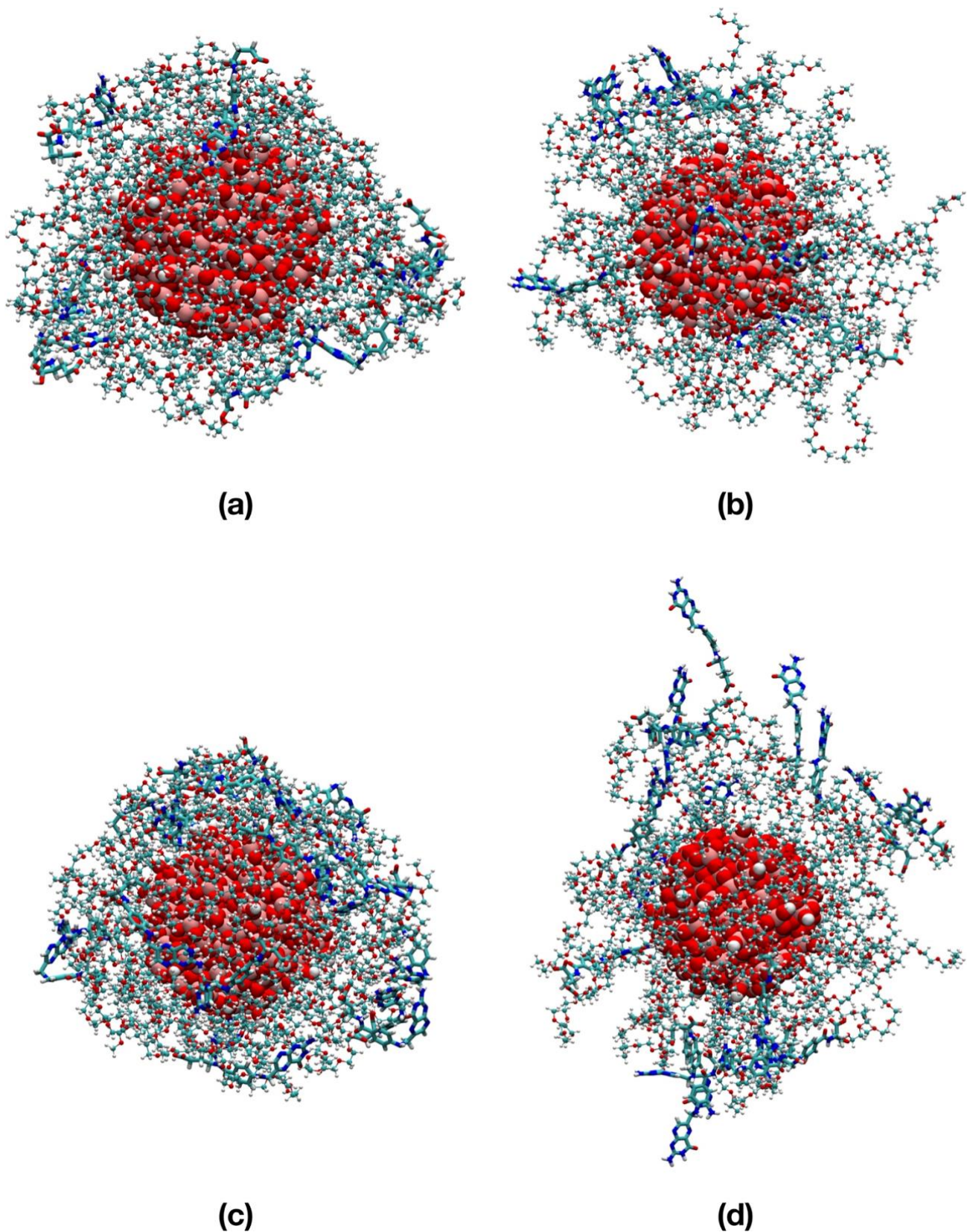

Fig. 9. Last snapshots from the 100 ns production simulation of TiO$_2$/PEG/10-FA system in vacuum (a) and in water (b) and TiO$_2$/PEG/20-FA system in vacuum (c) and in water (d). Titanium is shown

in pink, oxygen in red, carbon in cyan, nitrogen in blue and hydrogen in white. The water molecules are not shown for clarity.

In Fig. 10 we also report the rdf of the PEG and FA center of mass with respect to the six-fold coordinated Ti atom at the center of the nanoparticle, averaged over the 100 ns MD simulations, together with the total and LJ free energy profiles to transfer a FA molecule covalently attached to a PEG chain, from the NP surface towards the farthest distance compatible with the PEG chain length (see section 2.3 for more details).

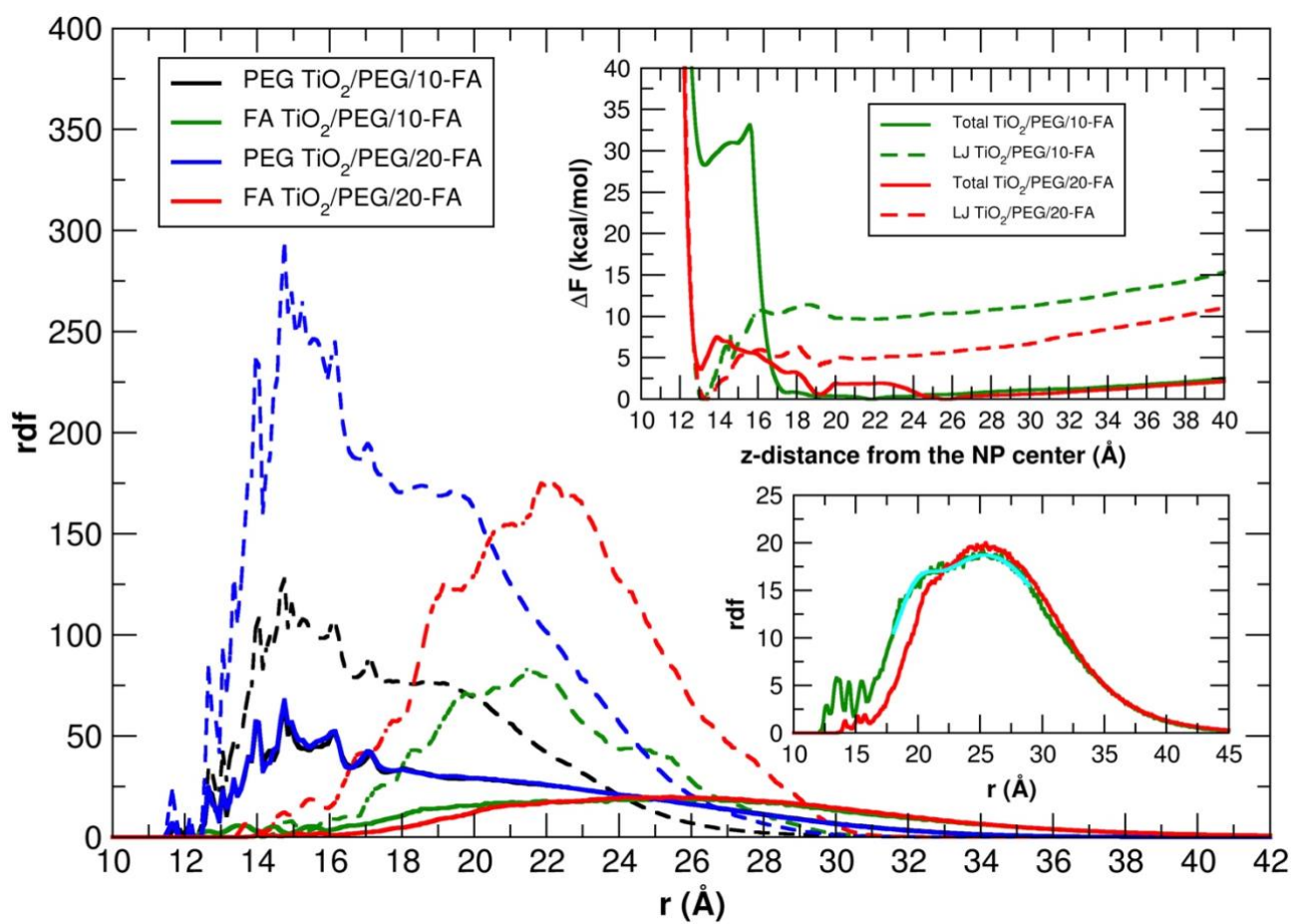

Fig. 10. Radial distribution function of the PEG and FA center of mass for the TiO$_2$/PEG/10-FA and TiO$_2$/PEG/20-FA models with respect to the six-fold coordinated Ti atom at the center of the nanoparticle, averaged over 100 ns MD simulations in vacuum (dashed lines) and in water (solid lines). Bottom-right inset: insight of the two FA rdf profiles in water, in cyan the double-gaussian function used to fit the FA rdf profile of the TiO$_2$/PEG/10-FA system in water is shown. Top-right inset: total and LJ free energy profiles for the diffusion of a FA molecule covalently attached to a PEG chain, from the NP surface towards the farthest distance compatible with the PEG chain length.

The distribution of the PEG molecules is similar in vacuum and in water for both the TiO$_2$/PEG/10-FA and TiO$_2$/PEG/20-FA models, with the maximum of the curves centered at about 15 Å from the Ti atom at the center of the NP. The trends of the FA ligands, instead, look different. In vacuum, the profiles present a principal peak and a few shoulders around it: for the TiO$_2$/PEG/10-FA and the TiO$_2$/PEG/20-FA systems, the maximum peaks are centered at about 21.5 Å and 22 Å from the NP center, respectively. In water, the two FA rdfs are broader: while the TiO$_2$/PEG/20-FA curve has a global maximum centered at 25.5 Å from the center of the NP (red plot in the bottom-right inset of Fig. 10), the TiO$_2$/PEG/10-FA rdf profile (in green) is characterized by a principal peak, centered at 25 Å from the NP center, and a shoulder, located at 21 Å, as obtained from a double-gaussian fitting of the curve (the fitting trend is shown in cyan). Moreover, there are several more ligands in the range of 10-15 Å in TiO$_2$/PEG/10-FA than in in TiO$_2$/PEG/20-FA. These results suggest that in TiO$_2$/PEG/20-FA the FA molecules are more exposed towards the bulk water phase, which nicely agrees with the experimental findings by Naghibi et al.[28] of maximized targeting activity for the corresponding weight ratio of 0.5, as mentioned above.

We can gain some valuable insight by comparing the FA rdf profile with that we obtained in another work for the cRGD peptide,[61] as mentioned above, in terms of ligand presentation and exposure on the same PEGylated TiO$_2$ NP surface. In Figure 11a, we overlay the FA rdf profile of the TiO$_2$/PEG/10-FA model (red line) with the one of TiO$_2$/PEG/10-cRGD (blue line), where 10 out of 50 PEG chains are conjugated with the targeting ligands. On the one hand, we observe a higher amount of FA ligands buried in the PEG inner region (12.5-25.0 Å) compared to cRGD ligands. On the other hand, we see that cRGD ligands remain more exposed in the PEG outer region than FA ligands (25.0-40.0 Å). These predictions, based on the MD simulations, remarkably agree with the experimental work by Valencia et al.,[62] who found that cRGD peptide stays more exposed on the PEGylated NP surface while FA tends to be more internalized into the polymeric NP, probably due to its higher hydrophobicity compared to cRGD.

In the top-right inset, the results of the free energy calculations are reported (refer to section 2.3 for computational details). The LJ profiles of both the TiO$_2$/PEG/10-FA and the TiO$_2$/PEG/20-FA systems (dashed-green and dashed-red lines, respectively) present a global minimum at z-distances of about 13 and 13.5 Å from the NP center, which correspond to the restrained, uncharged FA molecule (see section 2.3) interacting with the NP surface by short-range vdW forces. Moreover, for the TiO$_2$/PEG/20-FA system, the position of this minimum (dashed-red line) coincides with that of a local minimum in the total free energy curve (solid-red line), whose global minimum, however, is set at z-distance of about 25.5 Å, in good agreement with the position of the maximum (solid-red line) in the corresponding rdf profile (at about 25.5 Å). Differently, the total free energy profile of the TiO$_2$/PEG/10-FA system (solid-green line) is characterized by a first local minimum at a z-distance of about 13.5 Å from the NP center and another global minimum at about 22 Å, which well justifies the shoulder of the relative rdf curve (solid-green line, at about 21 Å).

To better understand this behavior, in Fig. S10 we report some representative snapshots of the TiO$_2$/PEG/10-FA system from the MD/ABF calculations, with the FA molecule restrained at different z-distances from the NP surface. In particular, we observe that the *p*-aminobenzoate ring changes its orientation from parallel (z-distance 10-15 Å, Fig. S10a) to perpendicular (z-distance 15-20 Å, Fig. S10b) with respect to the NP surface: we believe that this second configuration is associated to a higher free energy. For the sake of completeness, in Fig. S11 we also report the snapshots of the TiO$_2$/PEG/20-FA MD simulations. It is noteworthy to point out that the free energy profiles of Fig. 10 do not reach a plateau at increasing z-distance from the NP center: this is because the restrained FA molecule is covalently linked to a PEG chain, therefore pushing FA at high z-distances from the NP causes an increase of the free energy that comes from the constraint imposed by attached PEG chain, which in turn is covalently bound to the NP.

Finally, we analyze the dynamical properties of the TiO$_2$/PEG/FA nanoconjugates in an aqueous environment. In Fig. S12, the root-mean-square deviation (RMSD) plots of the NP+PEG+FA system's atomic positions along the 100 ns MD simulations, with respect to the 0 ns-reference atomic positions, are shown. The trends of RMSD vs time are similar: the slopes of the two curves have comparable magnitudes, which suggests that, according to our simulations, the two systems, i.e. the TiO$_2$/PEG/10-FA and the TiO$_2$/PEG/20-FA ones, diffuse in water with equivalent velocities. This is reasonable, since the two systems differ only for the number of the FA molecules, while the core of the nanosystem, i.e. the TiO$_2$ nanoparticle functionalized with the 50 PEG molecules, which mainly contributes to the mass of the whole structures, is the same.

To confirm these considerations, in Table S6 we report the estimated diffusion coefficients of the two TiO$_2$/PEG/FA systems in water, computed by the OLS method at MD trajectory intervals showing a linear dependence between MSD and time. Once again, as in section 3.2.3, the diffusion coefficient values are in the order of magnitude of $10^{-10}$ m$^2$/s. In addition, the diffusion results slightly slower when the FA density is higher.

**3.4. MD simulations in physiological environment**

This last section presents the MD results for the high-coverage TiO$_2$/n-FA and the TiO$_2$/PEG/n-FA systems under physiological conditions (0.15 M NaCl aqueous solution). Here, we aim at understanding how the introduction of some ionic strength affects the structural and dynamical properties of the nanoconjugates. For this purpose, we have performed the same structural analysis as previously done in Sections 3.2.1-3.3.2 for the TiO$_2$/52-FA-α, TiO$_2$/48-FA-γ, TiO$_2$/PEG/10-FA, and TiO$_2$/PEG/20-FA systems in pure water, over the last 100 ns of MD production phase, as detailed in Table S7 of the ESI. In the case of the TiO$_2$/52-FA-α and TiO$_2$/48-FA-γ models, no significant changes are observed, in particular, in the rdf profiles of FA in Fig. S13, suggesting that the distribution of FA molecules tethered on the NP surface is little affected by the presence of ions in solution (Table S7 vs Table S3). This was quite predictable considering the lower number of degrees of freedom for TiO$_2$/n-FA systems, where FA molecules are directly anchored to the NP surface, than for TiO$_2$/PEG/n-FA. In this respect, we would expect more significant changes for the PEGylated systems. Indeed, according to data in Table S7, we observe an enhancement in the FA-solution interactions, evidenced by a higher number of H-bonds and more favorable intermolecular interaction energies. On the other hand, the NP-FA interactions, which are already weak in pure water, become essentially null under physiological conditions. In Fig. 11b,c, rdf profiles for TiO$_2$/PEG/n-FA with n = 10 or 20, respectively, in pure water and in physiological conditions are superimposed. It is evident that the rdf profiles of FA are narrower and more shifted towards the bulk-water phase in the presence of NaCl in solution, where FA can interact with the ions, and, also, that this effect is more pronounced in the TiO$_2$/PEG/20-FA system. Moreover, under physiological conditions, the intensity of the FA rdf is null in the region closer to the NP surface for both systems with n = 10 and n = 20, whereas in the PEG inner region is lower compared to the simulations in pure water. Our findings are encouraging, as they suggest that in more realistic conditions (e.g. in a solution with ionic strength) these functionalized and PEGylated nanoconjugates would serve as more efficient nanomedical devices, being the FA moieties less interacting with the NP surface and more exposed to the extracellular fluid and, hence, to the folate receptors.

Finally, in Table S8 we report the estimated diffusion coefficients for the high-coverage TiO$_2$/n-FA and for the TiO$_2$/PEG/n-FA systems under physiological conditions. All the values are in the order of magnitude of $10^{-10}$ m$^2$/s, similar to the ones predicted in pure water (Table S4 and S6). Our results show that the presence of ions in solution has little impact on the diffusion of the nanoconjugates in solution.

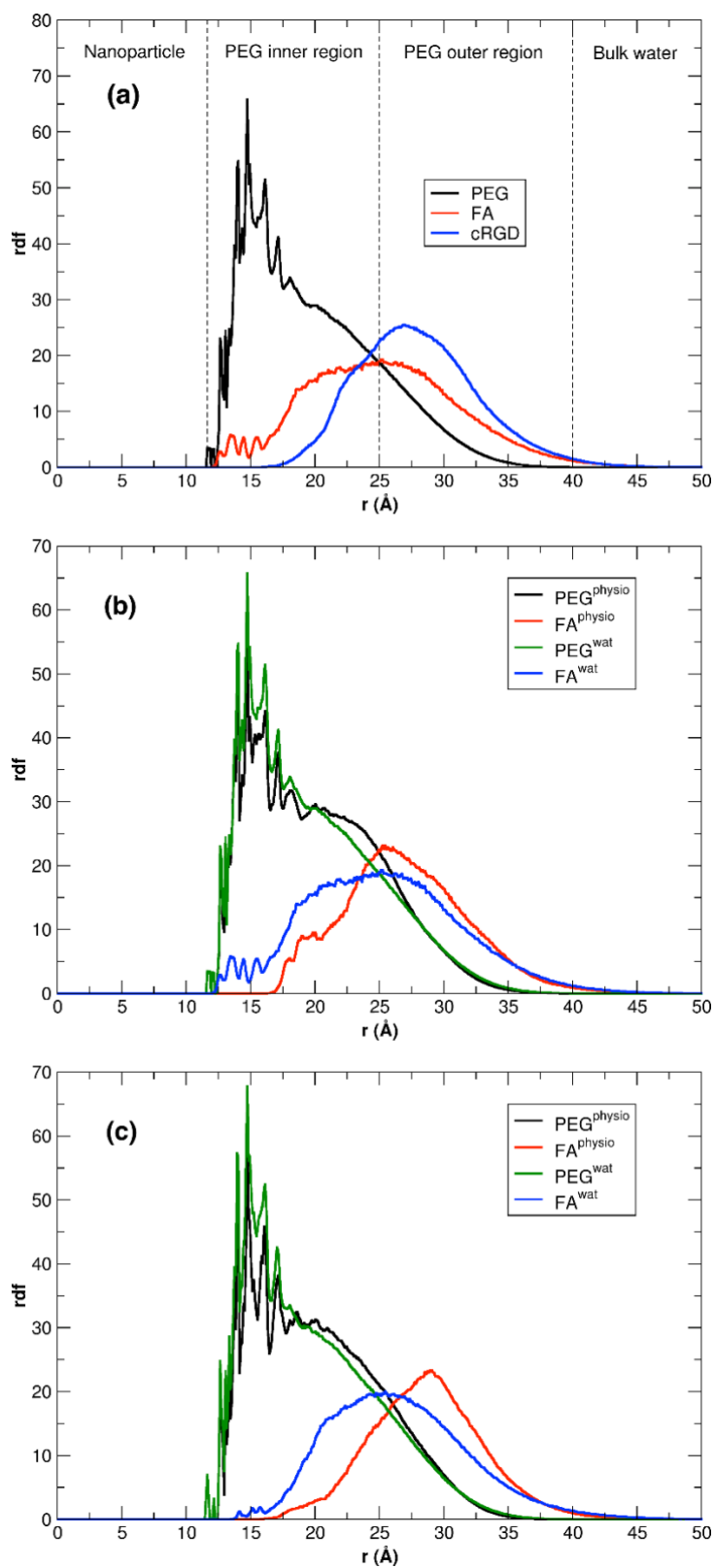

Fig. 11. (a) Comparison of FA and cRGD radial distribution function when conjugated to PEGylated TiO$_2$ NPs. PEG and FA profiles are the ones calculated in this work for TiO$_2$/PEG/10-FA model, while the cRGD profile was taken from our recent work[61] for the TiO$_2$/PEG system conjugated with the same number of targeting ligands (10 out of 50 PEG chains are conjugated with cRGD). For the sake of comparison, the cRGD profile is normalized in order to get the same integral as the FA profile. PEG inner and outer regions are separated by the point in which PEG and FA profiles cross, while bulk-water phase starts where PEG density goes to zero. (b,c) Comparison of the PEG and FA rdf profiles in water and under physiological conditions for TiO$_2$/PEG/10-FA (b) and TiO$_2$/PEG/20-FA (c) models.

## 4. Conclusions

In this work, we have simulated the dynamics of titanium dioxide nanoparticles, when these are functionalized with folic acid, which is an efficient targeting agent of several tumor cell lines. First, we have built, through quantum mechanical modeling at the DFTB level of theory, several TiO$_2$/n-FA systems, with increasing folic acid density and considering both the carboxylic groups of the ligand as anchoring groups. Our results reveal that the dissociative chelated and bidentate binding modes are energetically more advantageous and that the $\alpha$-COO$^-$ is the best anchoring carboxylate group. Moreover, the DFTB simulated annealing calculations confirmed the chemical stability of the nanoconjugates, when the effect of the temperature (500 K) is considered. Using the DFTB-optimized geometries as a starting point, we have performed long classical molecular dynamics simulations (100 ns) both in vacuum, pure water and under physiological conditions. In general, we have observed a strong interplay in vacuum between folic acid molecules and the NP surface through hydrophobic and electrostatic interactions, whereas in water there is a fair competition between ligand/NP and ligand/solvent interactions. We have also determined the dynamical properties of the nanoconjugates by computing their diffusion coefficients in water and observed a nice correlation with the total mass of the systems.

Due to the fact that the close contact and the large interaction of folic acid molecules with the NP surface could hamper its binding to the folate receptor, we have decided to design models where a spacer is introduced between the NP and the FA molecule. In particular we prepared two models where the NP is grafted by 50 PEG chains, among which some (10 or 20) were selected to be covalently bonded to folic acid molecules at the opposite end. Based on the classical MD simulations of these two models and on some free energy calculations we could prove that the introduction of the molecular spacer reduces the extent of the NP-folic acid interactions, which becomes even less pronounced under physiological conditions.

However, we have noticed that the folic acid content is a crucial parameter that highly affects the amount of the FA-FA intermolecular interactions but not much the diffusion coefficient of the nanosystems. According to our results and in agreement with the experiments,[21,28] an optimal fairly low density of FA molecules in the nanoconjugates (FA/TiO$_2$ weight ratio of about 0.2 for TiO$_2$/n-FA and of about 0.5 for TiO$_2$/PEG/n-FA) is expected to boost more efficiently their performance in targeting tumor cells, due to a more pronounced availability of the ligands for the folate receptor.

**Acknowledgments**

The authors are grateful to Lorenzo Ferraro for his technical support and to Enrico Bianchetti for useful suggestions. The project has received funding from the European Research Council (ERC) under the European Union's HORIZON2020 research and innovation programme (ERC Grant Agreement No [647020]) and from the University of Milano-Bicocca FAQC 2020 for the project "Photodynamic therapy for brain tumors by multifunctional particles using in situ Cerenkov and radioluminescence light".

53   P. Siani, E. Donadoni, L. Ferraro, F. Re and C. Di Valentin, *Biochimica et Biophysica Acta (BBA) - Biomembranes*, 2022, **1864**, 183763.

54   Z. Wu, X. Li, C. Hou and Y. Qian, *Journal of Chemical & Engineering Data*, 2010, **55**, 3958–3961.

55   M. Fernández, F. Javaid and V. Chudasama, *Chemical Science*, 2018, **9**, 790–810.

56   G. Fazio, D. Selli, L. Ferraro, G. Seifert and C. Di Valentin, *ACS Applied Materials & Interfaces*, 2018, **10**, 29943–29953.

57   C. Ronchi, M. Datteo, M. Kaviani, D. Selli and C. Di Valentin, *The Journal of Physical Chemistry C*, 2019, **123**, 10130–10144.

58   S. Mallakpour and B. Seyfi, *Materials Chemistry and Physics*, 2022, **281**, 125809.

59   D. H. Everett, *Basic Principles of Colloid Science*, Royal Society of Chemistry, London, 1988.

60   R. F. Domingos, N. Tufenkji and K. J. Wilkinson, *Environmental Science & Technology*, 2009, **43**, 1282–1286.

61   P. Siani, G. Frigerio, E. Donadoni and C. Di Valentin, *Journal of Colloid and Interface Science*, 2022, **627**, 126–141.

62   P. M. Valencia, M. H. Hanewich-Hollatz, W. Gao, F. Karim, R. Langer, R. Karnik and O. C. Farokhzad, *Biomaterials*, 2011, **32**, 6226–6233.

Electronic Supplementary Information

# Multi-scale modeling of folic acid-functionalized TiO$_2$ nanoparticles for active targeting of tumor cells

Edoardo Donadoni[a], Paulo Siani[a], Giulia Frigerio[a], Cristiana Di Valentin[a,b,*]

[a]Dipartimento di Scienza dei Materiali, Università di Milano-Bicocca
via R. Cozzi 55, 20125 Milano Italy
[b]BioNanoMedicine Center NANOMIB, University of Milano-Bicocca, Italy

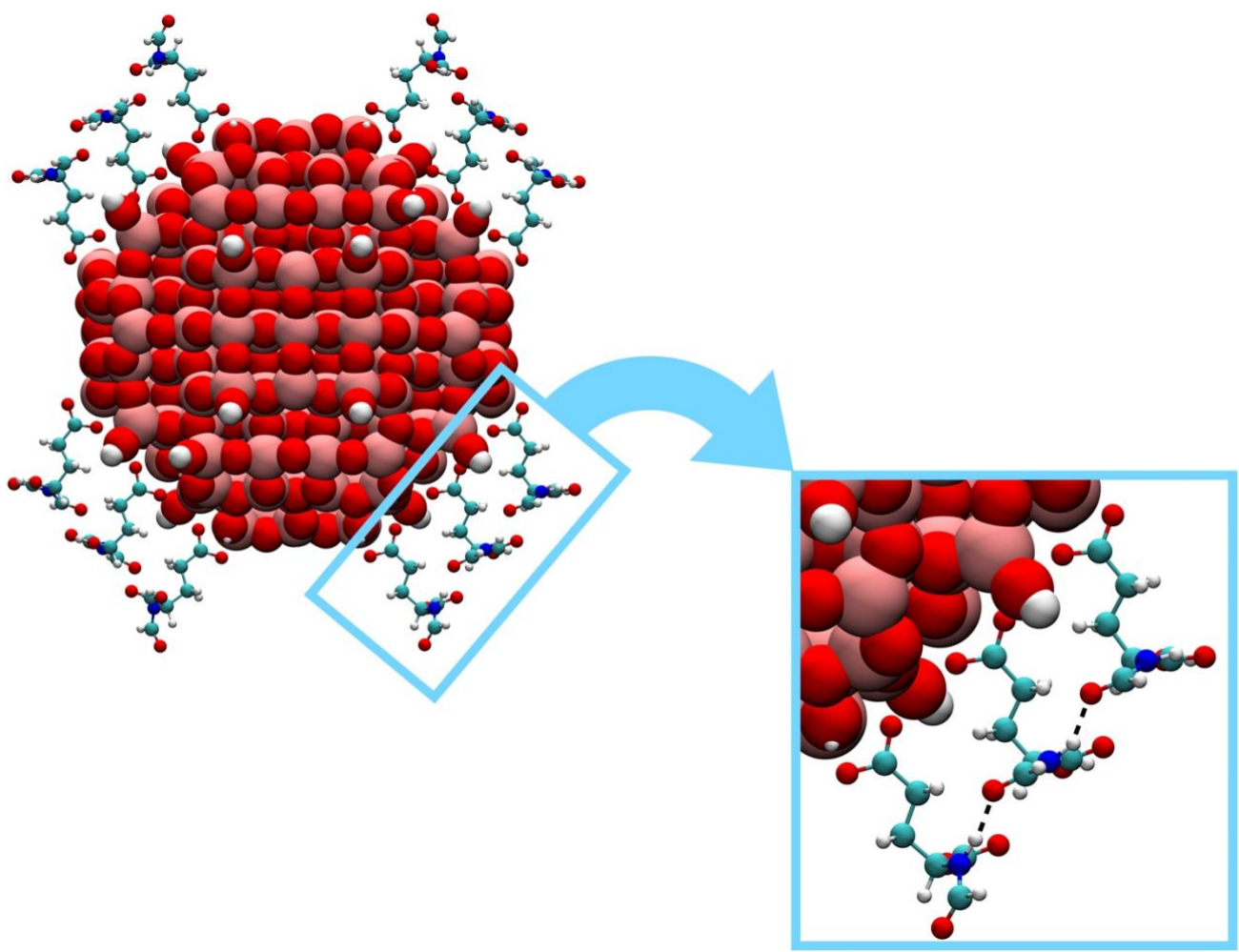

Fig. S1. Graphical representation of the TiO$_2$/8-FGA-$\gamma$ system and insight on the equatorial H-bonds network. Titanium is shown in pink, oxygen in red, carbon in cyan, nitrogen in blue and hydrogen in white.

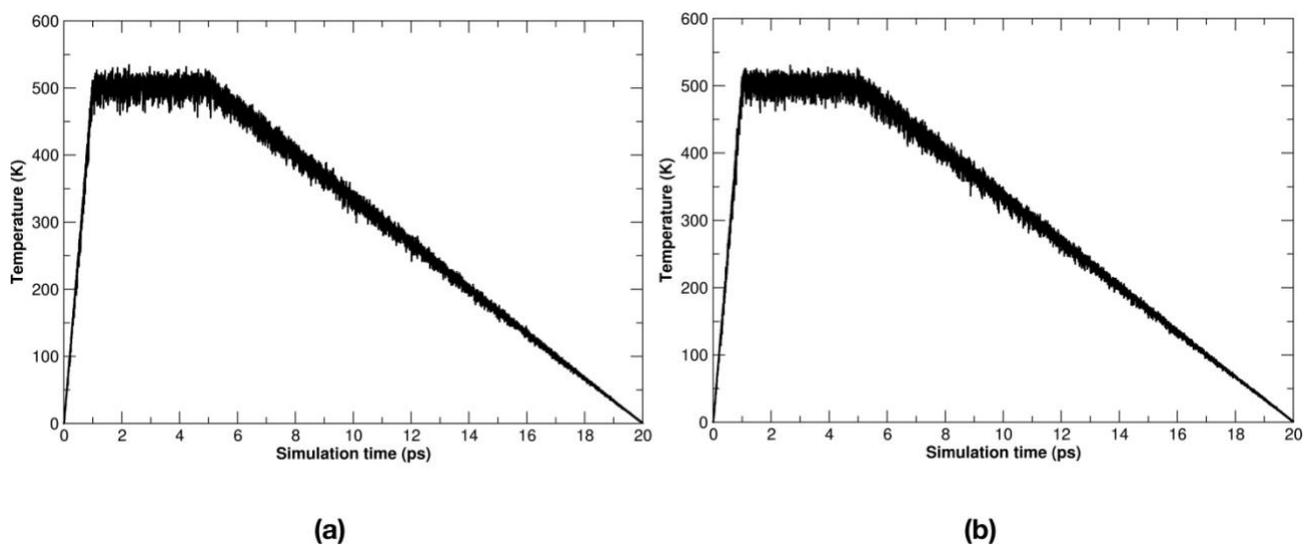

Fig. S2. Temperature profiles adopted for the DFTB-simulated annealing calculations for TiO$_2$/52-FGA-$\alpha$ (a) and TiO$_2$/48-FGA-$\gamma$ (b) systems.

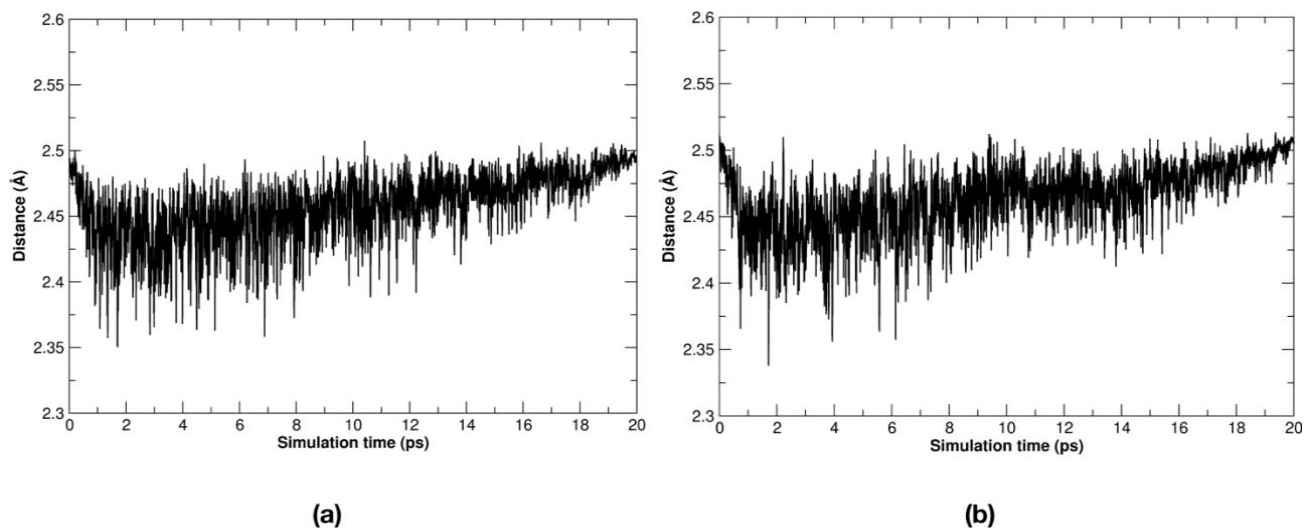

Fig. S3. Time evolution of the average distance between the C atom of FGA carboxylic groups anchoring the NP and the coordinated Ti atom of the NP, along the DFTB-simulated annealing calculations, for TiO$_2$/52-FGA-$\alpha$ (a) and TiO$_2$/48-FGA-$\gamma$ (b) systems.

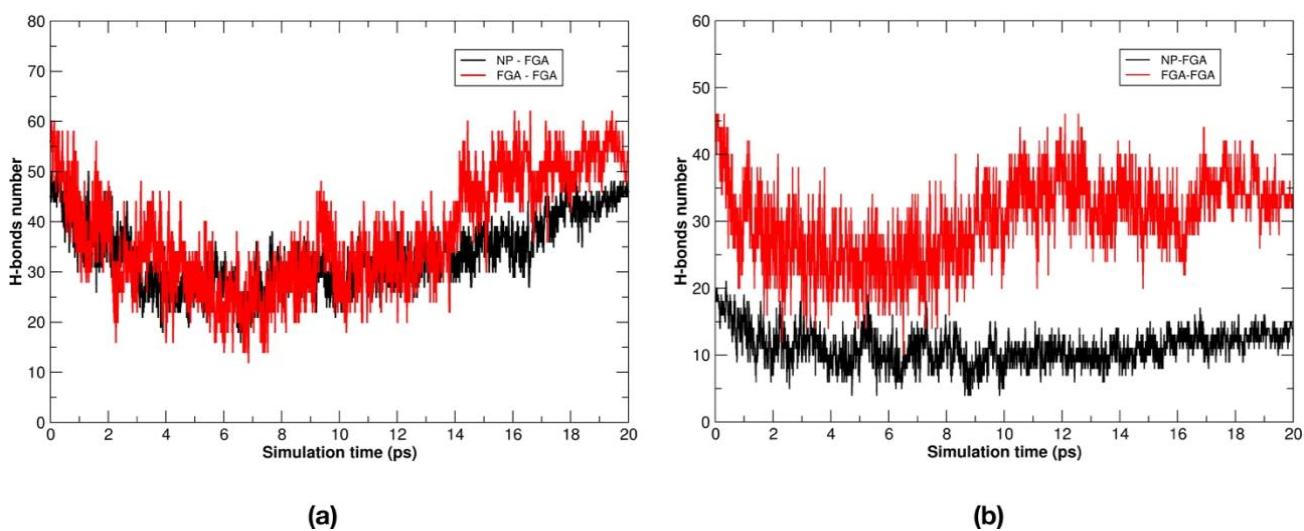

Fig. S4. Time evolution of the NP-FGA and FGA-FGA hydrogen bonds number, along the DFTB-simulated annealing calculations, for TiO$_2$/52-FGA-$\alpha$ (a) and TiO$_2$/48-FGA-$\gamma$ (b) systems.

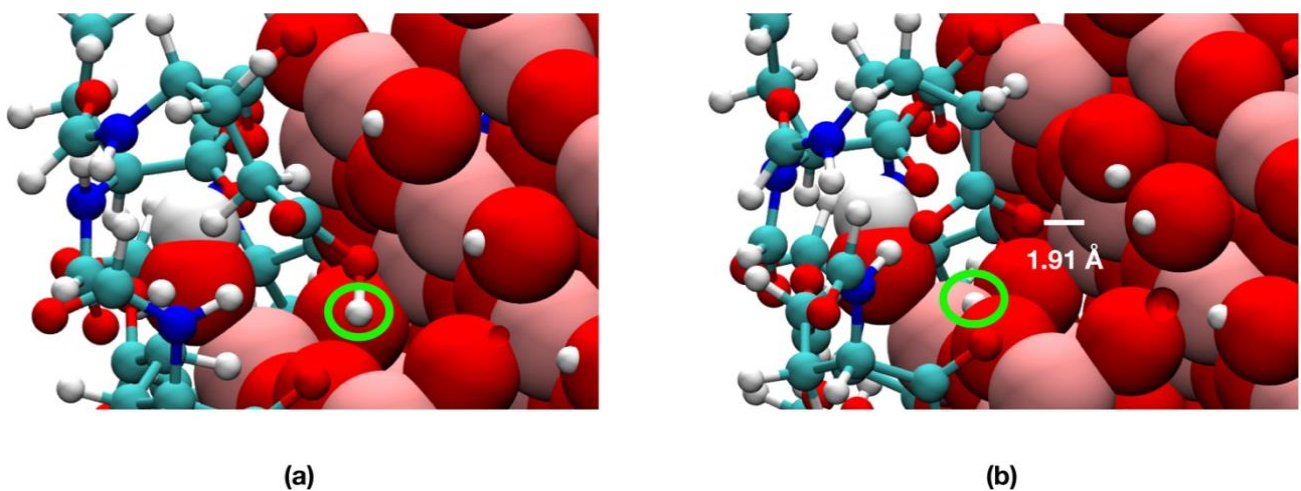

Fig. S5. Proton transfer from the free carboxylic group of one FGA molecule (a) to a near O$_{2c}$ atom of the NP (b), with the formation of one NP-FGA Ti-O bond in a monodentate configuration, during the DFTB-simulated annealing for the TiO$_2$/52-FGA-$\alpha$ system. Titanium is shown in pink, oxygen in red, carbon in cyan, nitrogen in blue and hydrogen in white.

Table S1. Average distances, hydrogen bonds number and non-bonding (vdW and electrostatic) interaction energies for the 100 ns production simulations of the TiO$_2$/1-FA-$\alpha$ and TiO$_2$/1-FA-$\gamma$ in vacuum and in water.

| **Indicator** | **TiO$_2$/1-FA-$\alpha$ vacuum** | **TiO$_2$/1-FA-$\gamma$ vacuum** | **TiO$_2$/1-FA-$\alpha$ water** | **TiO$_2$/1-FA-$\gamma$ water** |
|---|---|---|---|---|
| | **Distances (Å)** | | | |
| d N$^{FA}$-NP$^{center}$ | 15.5 ($\pm$ 0.1) | 13.5 ($\pm$ 0.1) | 18 ($\pm$ 4) | 16 ($\pm$ 1) |
| d com$^{FA}$-NP$^{surface}$ | 5.3 ($\pm$ 0.1) | 5.5 ($\pm$ 0.1) | 6 ($\pm$ 1) | 6 ($\pm$ 1) |
| | **Hydrogen bonds number** | | | |
| NP-FA | 0 | 0.6 ($\pm$ 0.7) | 0.1 ($\pm$ 0,3) | 0.04 ($\pm$ 0.21) |
| FA-wat | - | - | 9 ($\pm$ 2) | 9 ($\pm$ 2) |
| NP-wat | - | - | 39 ($\pm$ 4) | 39 ($\pm$ 5) |
| | **Non-bonding interaction energies (kcal/mol)** | | | |
| vdW NP-FA | -22 ($\pm$ 1) | -27 ($\pm$ 5) | -19 ($\pm$ 6) | -21 ($\pm$ 5) |
| ele NP-FA | -38 ($\pm$ 2) | -19 ($\pm$ 3) | -5 ($\pm$ 4) | -8 ($\pm$ 4) |
| vdW FA-wat | - | - | -18 ($\pm$ 5) | -17 ($\pm$ 5) |
| ele FA-wat | - | - | -95 ($\pm$ 10) | -97 ($\pm$ 10) |
| vdW NP-wat | - | - | -443 ($\pm$ 13) | -456 ($\pm$ 12) |
| ele NP-wat | - | - | -523 ($\pm$ 21) | -527 ($\pm$ 22) |

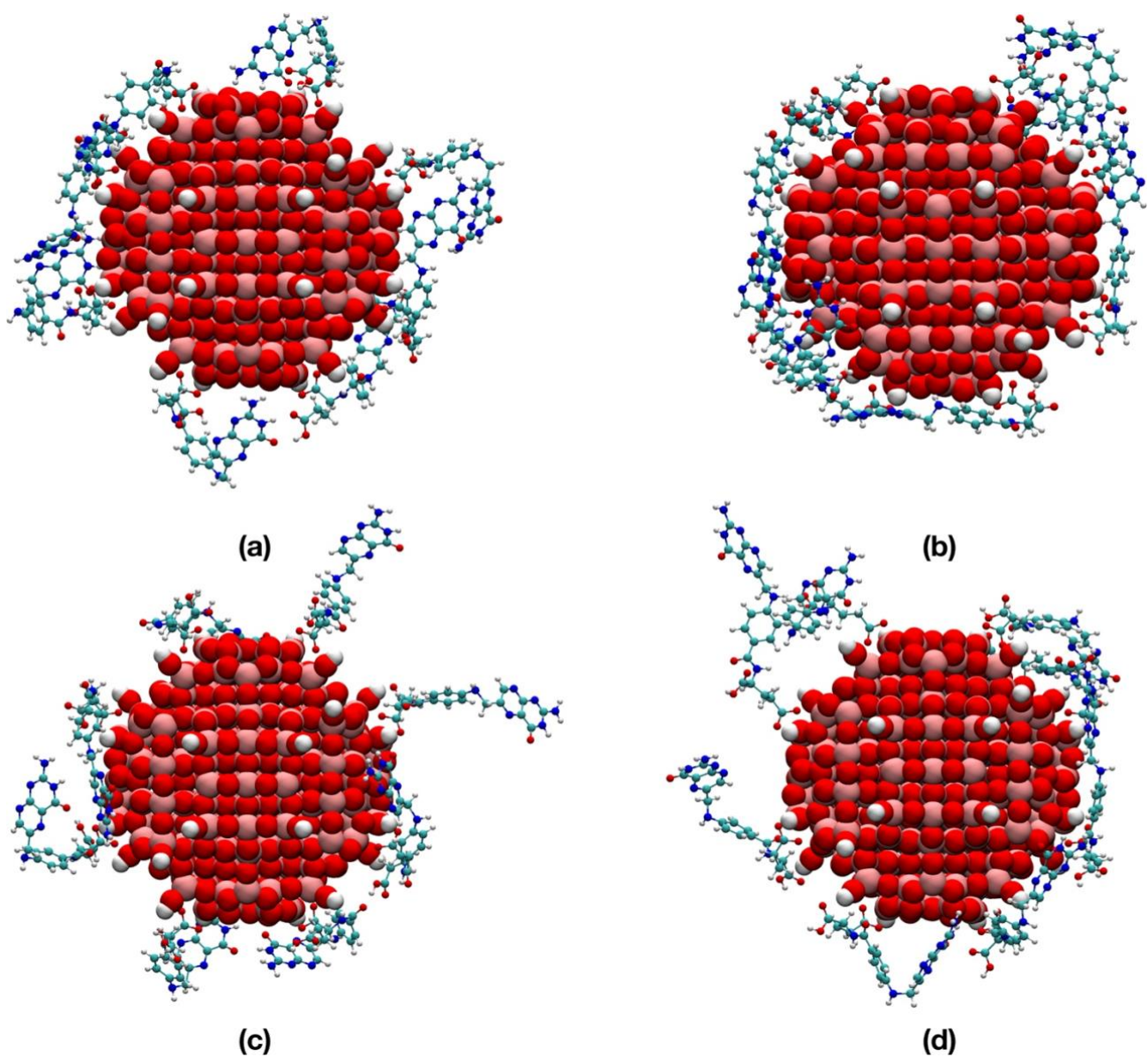

Fig. S6. Last snapshots from the 100 ns production simulations of TiO$_2$/8-FA-$\alpha$ in vacuum (a) and in water (c) and TiO$_2$/8-FA-$\gamma$ in vacuum (b) and in water (d). Titanium is shown in pink, oxygen in red, carbon in cyan, nitrogen in blue and hydrogen in white. The water molecules are not shown for clarity.

Table S2. Average distances, hydrogen bonds number and non-bonding (vdW and electrostatic) interaction energies for the 100 ns production simulations of the TiO$_2$/8-FA-$\alpha$ and TiO$_2$/8-FA-$\gamma$ in vacuum and in water.

| Indicator | TiO$_2$/8-FA-$\alpha$ vacuum | TiO$_2$/8-FA-$\gamma$ vacuum | TiO$_2$/8-FA-$\alpha$ water | TiO$_2$/8-FA-$\gamma$ water |
|---|---|---|---|---|
| **Distances (Å)** | | | | |
| d N$^{FA}$-NP$^{center}$ | 16 ($\pm$ 2) | 16 ($\pm$ 2) | 20 ($\pm$ 8) | 20 ($\pm$ 11) |
| d com$^{FA}$-NP$^{surface}$ | 6 ($\pm$ 1) | 6 ($\pm$ 1) | 7 ($\pm$ 1) | 8 ($\pm$ 4) |
| **Hydrogen bonds number** | | | | |
| NP-FA | 4 ($\pm$ 1) | 3 ($\pm$ 1) | 0.2 ($\pm$ 0,4) | 0.9 ($\pm$ 0,9) |
| FA-FA | 5 ($\pm$ 3) | 5 ($\pm$ 2) | 0.3 ($\pm$ 0,8) | 0.9 ($\pm$ 1.2) |
| FA-wat | - | - | 72 ($\pm$ 5) | 69 ($\pm$ 6) |
| NP-wat | - | - | 32 ($\pm$ 4) | 34 ($\pm$ 4) |
| **Non-bonding interaction energies (kcal/mol)** | | | | |
| vdW NP-FA | -185 ($\pm$ 9) | -209 ($\pm$ 5) | -176 ($\pm$ 11) | -115 ($\pm$ 24) |
| ele NP-FA | -107 ($\pm$ 9) | -110 ($\pm$ 8) | -28 ($\pm$ 7) | -31 ($\pm$ 11) |
| vdW FA-FA | -16 ($\pm$ 3) | -24 ($\pm$ 4) | -3 ($\pm$ 2) | -24 ($\pm$ 5) |
| ele FA-FA | -78 ($\pm$ 10) | -61 ($\pm$ 7) | -3 ($\pm$ 5) | -11 ($\pm$ 9) |
| vdW FA-wat | - | - | -157 ($\pm$ 13) | -149 ($\pm$ 14) |
| ele FA-wat | - | - | -766 ($\pm$ 28) | -756 ($\pm$ 33) |
| vdW NP-wat | - | - | -420 ($\pm$ 13) | -456 ($\pm$ 18) |
| ele NP-wat | - | - | -515 ($\pm$ 23) | -541 ($\pm$ 27) |

Table S3. Average distances, hydrogen bonds number and non-bonding (vdW and electrostatic) interaction energies for the 100 ns production simulations of the TiO$_2$/52-FA-$\alpha$ and TiO$_2$/48-FA-$\gamma$ systems in vacuum and in water.

| Indicator | TiO$_2$/52-FA-$\alpha$ vacuum | TiO$_2$/48-FA-$\gamma$ vacuum | TiO$_2$/52-FA-$\alpha$ water | TiO$_2$/48-FA-$\gamma$ water |
|---|---|---|---|---|
| **Distances (Å)** | | | | |
| d N$^{FA}$-NP$^{center}$ | 21 ($\pm$ 3) | 20 ($\pm$ 3) | 24 ($\pm$ 2) | 24 ($\pm$ 3) |
| d com$^{FA}$-NP$^{surface}$ | 8 ($\pm$ 3) | 8 ($\pm$ 2) | 8 ($\pm$ 1) | 9 ($\pm$ 2) |
| **Hydrogen bonds number** | | | | |
| NP-FA | 5 ($\pm$ 1) | 5 ($\pm$ 1) | 4 ($\pm$ 1) | 2 ($\pm$ 1) |
| FA-FA | 86 ($\pm$ 9) | 73 ($\pm$ 9) | 23 ($\pm$ 6) | 18 ($\pm$ 6) |
| FA-wat | - | - | 392 ($\pm$ 13) | 376 ($\pm$ 14) |
| NP-wat | - | - | 11 ($\pm$ 3) | 22 ($\pm$ 3) |
| **Non-bonding interaction energies (kcal/mol)** | | | | |
| vdW NP-FA | -612 ($\pm$ 10) | -535 ($\pm$ 15) | -610 ($\pm$ 9) | -424 ($\pm$ 19) |
| ele NP-FA | -167 ($\pm$ 14) | -180 ($\pm$ 13) | -128 ($\pm$ 12) | -86 ($\pm$ 12) |
| vdW FA-FA | -536 ($\pm$ 17) | -464 ($\pm$ 16) | -494 ($\pm$ 19) | -443 ($\pm$ 19) |
| ele FA-FA | -1308 ($\pm$ 37) | -1241 ($\pm$ 38) | -276 ($\pm$ 50) | -256 ($\pm$ 45) |
| vdW FA-wat | - | - | -643 ($\pm$ 34) | -652 ($\pm$ 34) |
| ele FA-wat | - | - | -4308 ($\pm$ 97) | -4110 ($\pm$ 106) |
| vdW NP-wat | - | - | -130 ($\pm$ 9) | -246 ($\pm$ 16) |
| ele NP-wat | - | - | -333 ($\pm$ 19) | -541 ($\pm$ 24) |

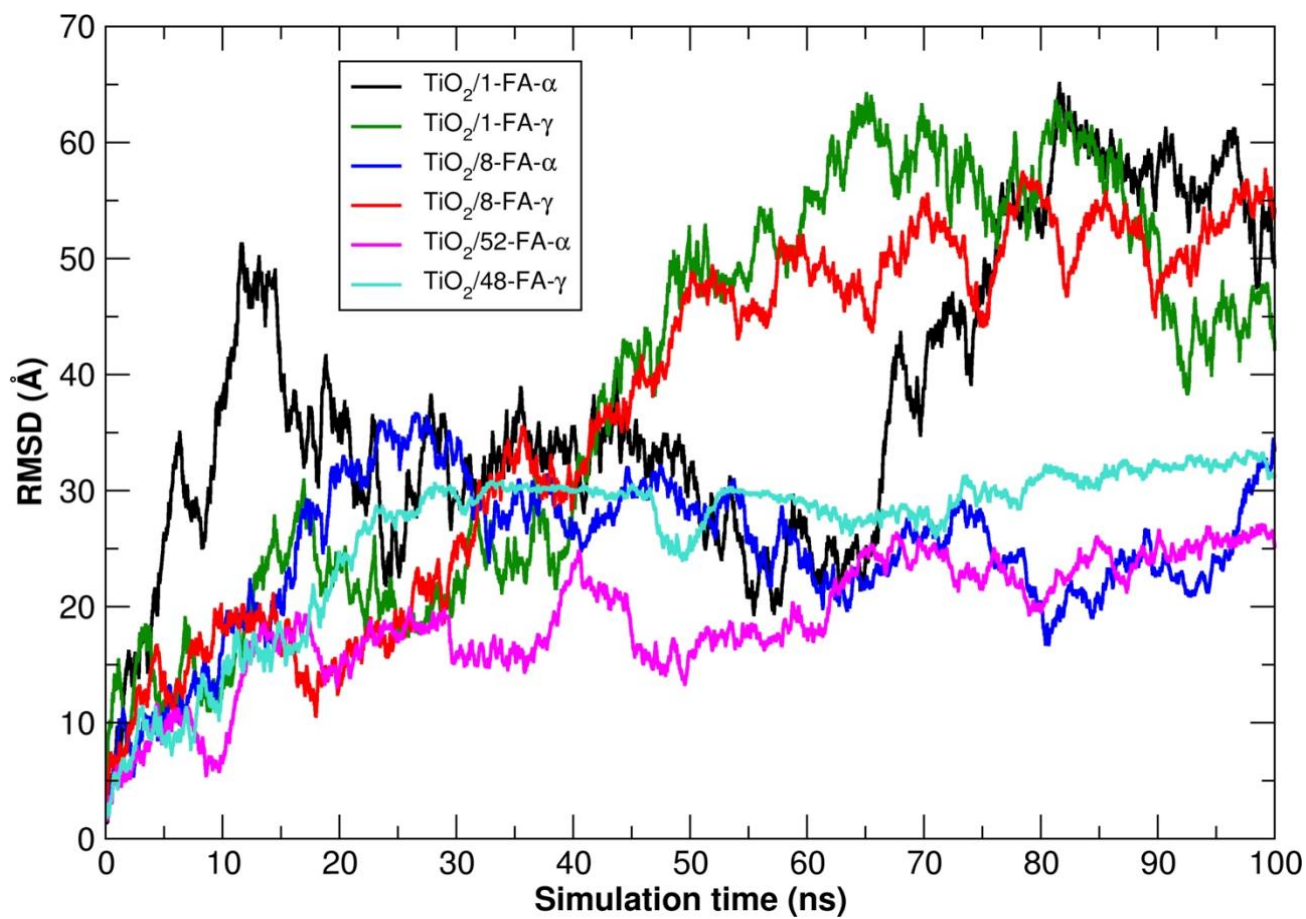

Fig. S7. Root-mean-square deviation of the NP+FA atomic positions along the 100 ns MD simulations, with respect to the 0 ns-reference atomic positions, for all the TiO$_2$/n-FA systems in water.

Table S4. Estimated NP+FA diffusion coefficients for the 100 ns production simulations of all the TiO$_2$/n-FA-$\alpha$ systems in water.

| System | D/10$^{-10}$ (m$^2$/s) |
|---|---|
| TiO$_2$/1-FA-$\alpha$ | 7.7 ($\pm$ 0.3) |
| TiO$_2$/8-FA-$\alpha$ | 2.06 ($\pm$ 0.06) |
| TiO$_2$/52-FA-$\alpha$ | 1.70 ($\pm$ 0.07) |

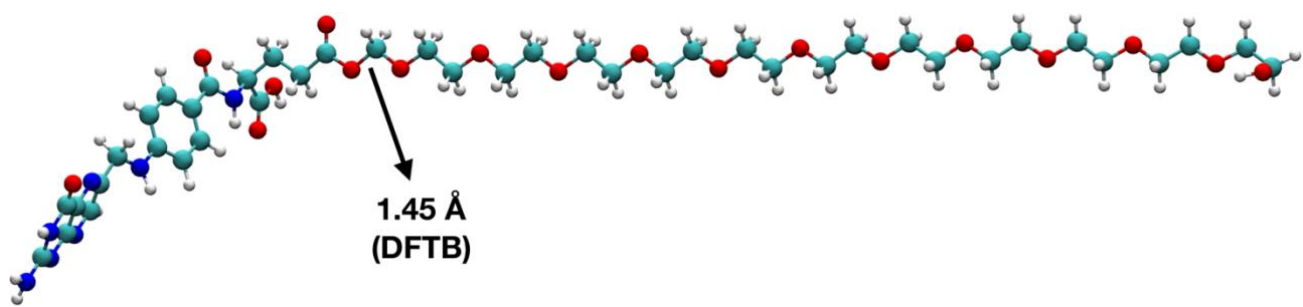

Fig. S8. DFTB-optimized geometry of a PEG-FA chain. Oxygen is shown in pink, carbon in cyan, nitrogen in blue and hydrogen in white.

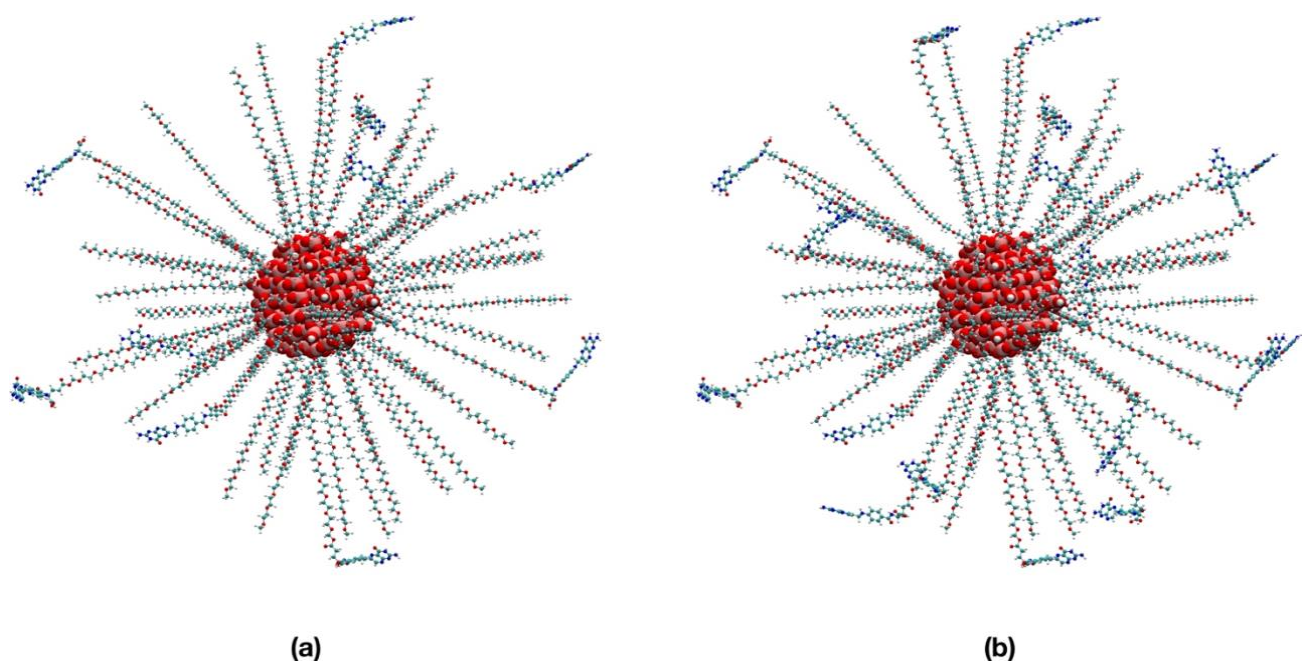

Fig. S9. DFTB-optimized geometries of the TiO$_2$/PEG/10-FA (a) and the TiO$_2$/PEG/20-FA (b) systems. Titanium is shown in pink, oxygen in red, carbon in cyan, nitrogen in blue and hydrogen in white.

Table S5. Average distances, hydrogen bonds number and non-bonding (vdW and electrostatic) interaction energies for the 100 ns production simulations of the TiO$_2$/PEG/10-FA and TiO$_2$/PEG/20-FA systems, in vacuum and in water.

| Indicator | TiO$_2$/PEG/10-FA vacuum | TiO$_2$/PEG/10-FA water | TiO$_2$/PEG/20-FA vacuum | TiO$_2$/PEG/20-FA water |
|---|---|---|---|---|
| **Distances (Å)** | | | | |
| d N$^{FA}$-NP$^{center}$ | 21 (± 2) | 31 (± 5) | 22 (± 3) | 32 (± 3) |
| d com$^{FA}$-NP$^{surface}$ | 12 (± 3) | 17 (± 2) | 13 (± 3) | 17 (± 2) |
| $\langle h^2 \rangle^{\frac{1}{2}\,PEG}$ | 13 (± 4) | 18 (± 1) | 13 (± 3) | 18 (± 1) |
| $\langle h^2 \rangle^{\frac{1}{2}\,PEG-FA}$ | 11 (± 4) | 16 (± 2) | 12 (± 3) | 17 (± 1) |
| MDFS$^{PEG}$ | 9 (± 3) | 14 (± 1) | 9 (± 2) | 14 (± 1) |
| MDFS$^{PEG-FA}$ | 11 (± 3) | 13 (± 2) | 9 (± 3) | 13 (± 1) |
| R$_g^{PEG}$ | 5.3 (± 0.1) | 6.7 (± 0.1) | 5.3 (± 0.1) | 6.8 (± 0.1) |
| R$_g^{PEG-FA}$ | 5.7 (± 0.1) | 7.6 (± 0.1) | 6.8 (± 0.1) | 8.8 (± 0.1) |
| **Hydrogen bonds number** | | | | |
| NP-FA | 0 | 0.7 (± 0.5) | 0.5 (± 0.5) | 0.3 (± 0.6) |
| NP-PEG | 0 | 0 | 0 | 0 |
| PEG-PEG | 3 (± 2) | 2 (± 1) | 1 (± 1) | 2 (± 1) |
| PEG-FA | 18 (± 3) | 2 (± 1) | 32 (± 4) | 5 (± 2) |
| FA-FA | 6 (± 3) | 1 (± 1) | 16 (± 4) | 2 (± 2) |
| FA-wat | - | 90 (± 6) | - | 178 (± 9) |
| PEG-wat | - | 308 (± 13) | - | 298 (± 13) |
| NP-wat | - | 59 (± 4) | - | 58 (± 3) |
| **Non-bonding interaction energies (kcal/mol)** | | | | |
| vdW NP-FA | -89 (± 7) | -12 (± 3) | -17 (± 1) | -5 (± 5) |
| ele NP-FA | -53 (± 5) | -18 (± 3) | -13 (± 3) | -6 (± 8) |
| vdW NP-PEG | -478 (± 13) | -274 (± 20) | -525 (±12) | -276 (± 18) |
| ele NP-PEG | -269 (± 10) | -99 (± 10) | -303 (± 11) | -114 (± 11) |
| vdW PEG-PEG | -820 (± 17) | -382 (± 22) | -808 (± 17) | -396 (± 24) |
| ele PEG-PEG | -517 (± 27) | 10 (± 13) | -169 (± 19) | 5 (± 15) |
| vdW PEG-FA | -335 (± 12) | -94 (± 20) | -420 (± 18) | -180 (± 33) |
| ele PEG-FA | -93 (± 17) | -46 (± 21) | -558 (± 31) | -98 (± 29) |
| vdW FA-FA | -6 (± 2) | -10 (± 9) | -60 (± 7) | -47 (± 16) |
| ele FA-FA | 0.4 (± 1.6) | -6 (± 6) | -205 (± 14) | -39 (± 18) |
| vdW FA-wat | - | -151 (± 24) | - | -298 (± 32) |
| ele FA-wat | - | -1041 (± 50) | - | -2059 (± 71) |
| vdW PEG-wat | - | -1490 (± 45) | - | -1421 (± 47) |
| ele PEG-wat | - | -3906 (± 88) | - | -3755 (± 93) |
| vdW NP-wat | - | -94 (± 19) | - | -98 (± 17) |
| ele NP-wat | - | -978 (± 37) | - | -950 (± 37) |

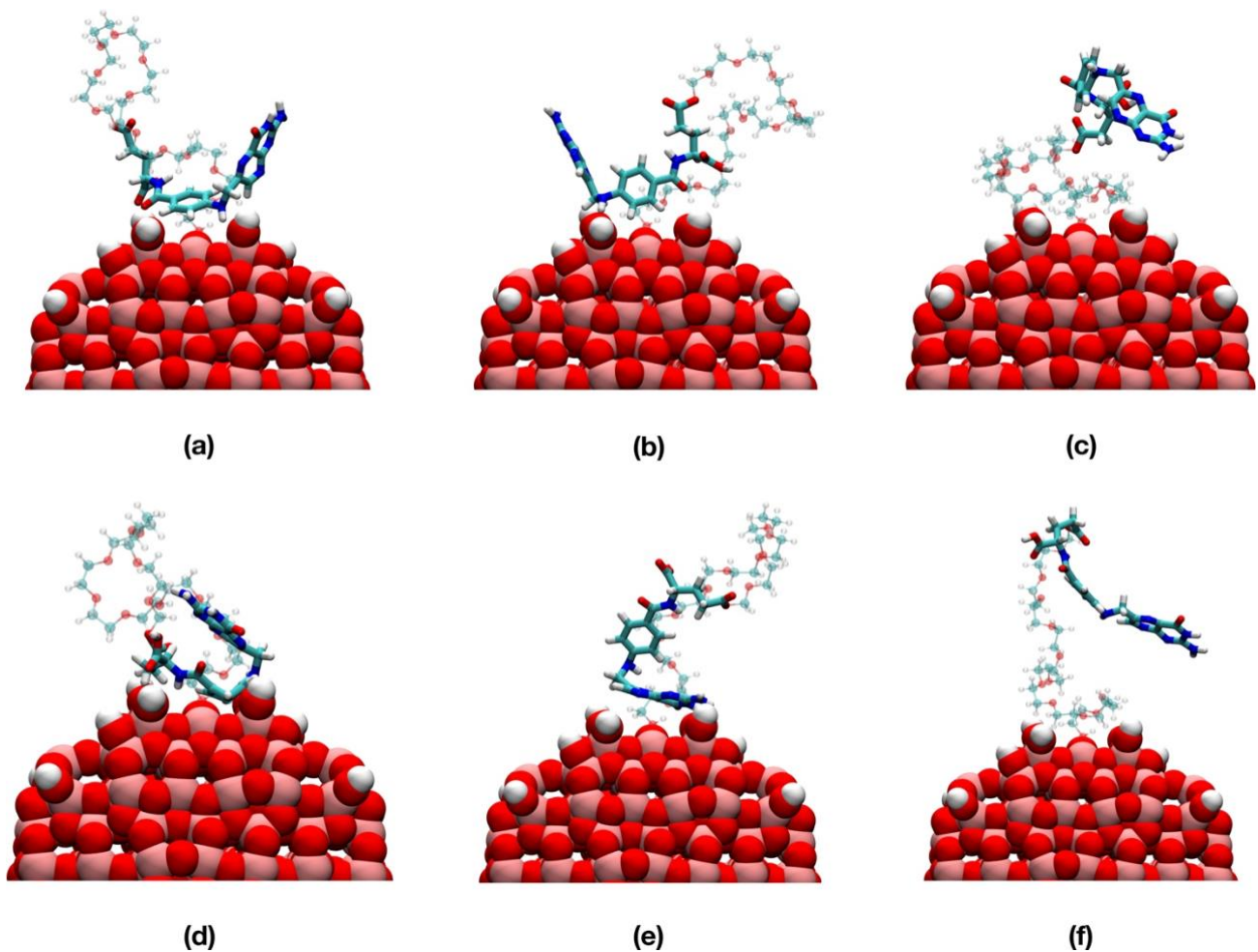

Fig. S10. Last snapshots from the 150 ns production ABF simulations of TiO$_2$/PEG/10-FA in water with a FA molecule (opaque) restrained respectively between 10-15 Å (a), 15-20 Å (b) and 20-25 Å (c) from the NP center with its relative PEG chain (transparent) and the ABF simulations with the same FA molecule uncharged and restrained respectively between 10-15 Å (d), 15-20 Å (e) and 20-25 Å (f) from the NP center. Titanium is shown in pink, oxygen in red, carbon in cyan, nitrogen in blue and hydrogen in white. The other PEG and FA molecules and the water molecules are not shown for clarity.

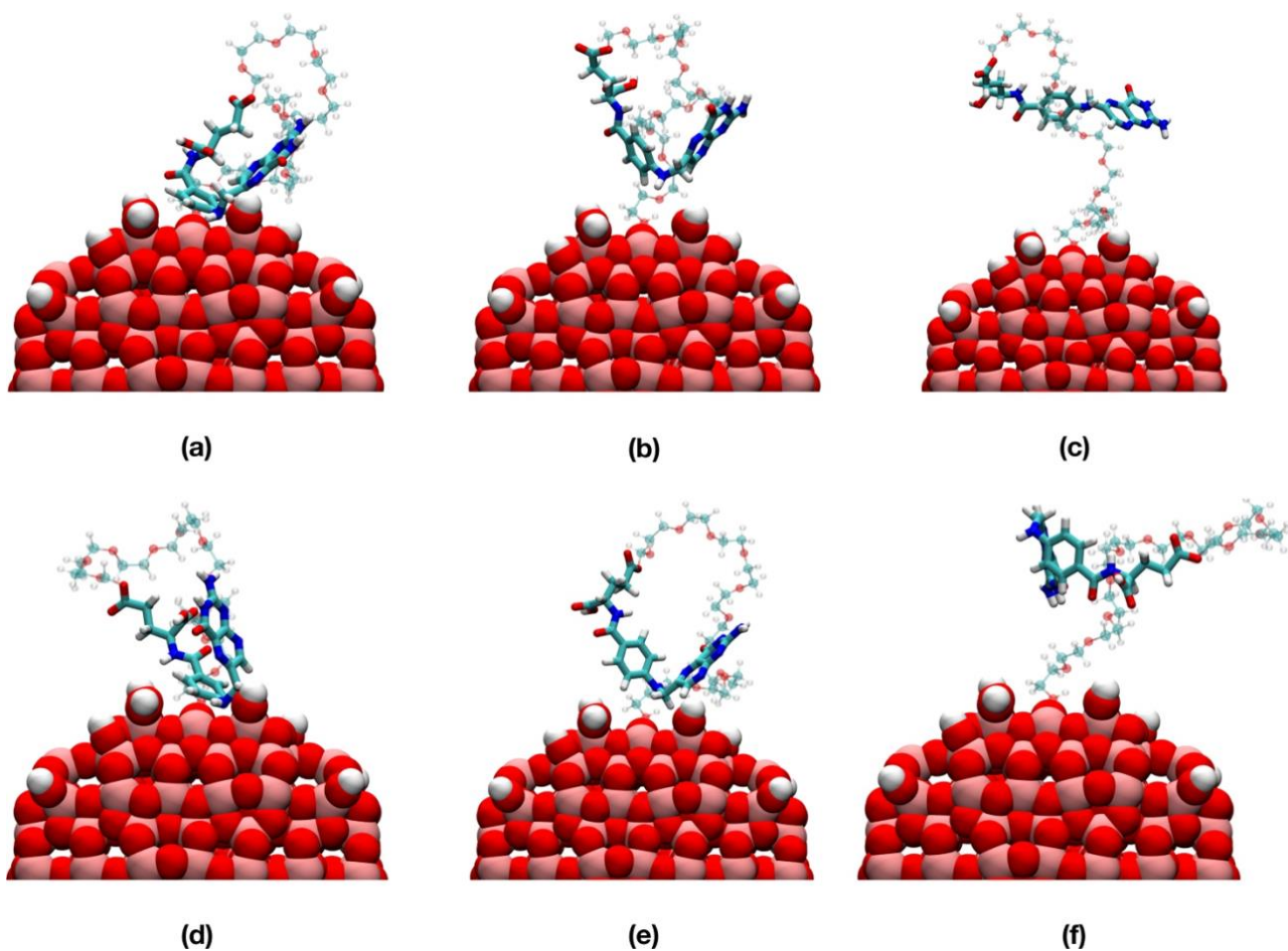

Fig. S11. Last snapshots from the 150 ns production ABF simulations of TiO$_2$/PEG/20-FA in water with a FA molecule (opaque) restrained respectively between 10-15 Å (a), 15-20 Å (b) and 20-25 Å (c) from the NP center with its relative PEG chain (transparent) and the ABF simulations with the same FA molecule uncharged and restrained respectively between 10-15 Å (d), 15-20 Å (e) and 20-25 Å (f) from the NP center. Titanium is shown in pink, oxygen in red, carbon in cyan, nitrogen in blue and hydrogen in white. The other PEG and FA molecules and the water molecules are not shown for clarity.

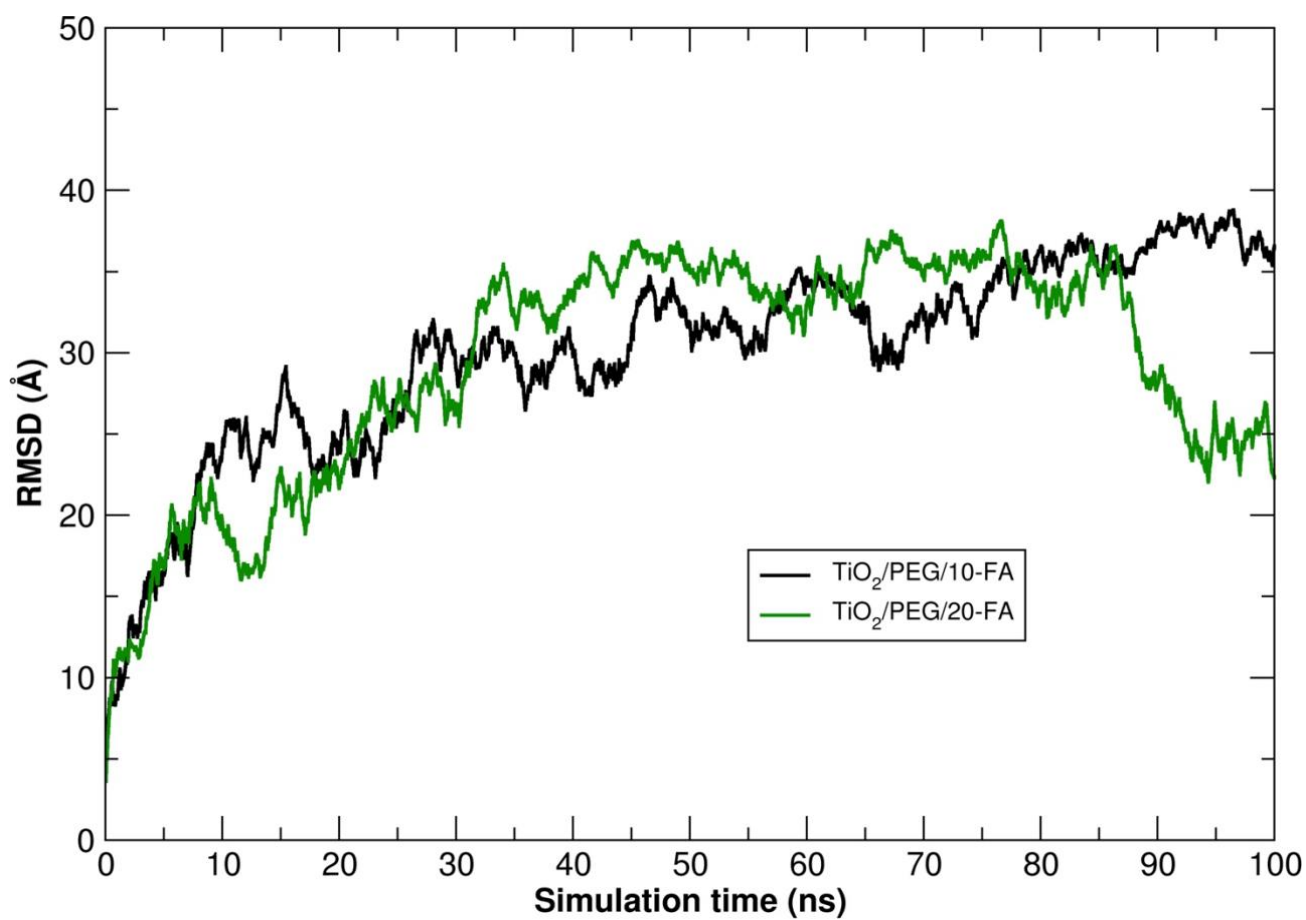

Fig. S12. Root-mean-square deviation of the NP+PEG+FA atomic positions along the 100 ns MD simulations, with respect to the 0 ns-reference atomic positions, for the TiO$_2$/PEG/10-FA and TiO$_2$/PEG/20-FA systems in water.

Table S6. Estimated NP+PEG+FA diffusion coefficients for the 100 ns production simulations of the TiO$_2$/PEG/10-FA and TiO$_2$/PEG/20-FA systems in water.

| System | D/$10^{-10}$ (m$^2$/s) |
|---|---|
| TiO$_2$/PEG/10-FA | 2.5 ($\pm$ 0.4) |
| TiO$_2$/PEG/20-FA | 2.32 ($\pm$ 0.09) |

Table S7. Average distances, number of hydrogen bonds and non-bonding (vdW and electrostatic) interaction energies for the last 100 ns of MD production phase of the TiO$_2$/52-FA-$\alpha$, TiO$_2$/48-FA-$\gamma$, TiO$_2$/PEG/10-FA and TiO$_2$/PEG/20-FA systems in physiological environment.

| Indicator | TiO$_2$/52-FA-$\alpha$ | TiO$_2$/48-FA-$\gamma$ | TiO$_2$/PEG/10-FA | TiO$_2$/PEG/20-FA |
|---|---|---|---|---|
| | Distances (Å) | | | |
| d N$^{FA}$-NP$^{center}$ | 24 (± 2) | 24 (± 3) | 30 (± 3) | 32 (± 2) |
| d com$^{FA}$-NP$^{surface}$ | 8 (± 1) | 9 (± 2) | 17 (± 3) | 18 (± 2) |
| $<h^2>^{½\ PEG}$ | - | - | 16 (± 2) | 17 (± 2) |
| $<h^2>^{½\ PEG-FA}$ | - | - | 16 (± 3) | 16 (± 2) |
| MDFS$^{PEG}$ | - | - | 13 (± 1) | 13 (± 1) |
| MDFS$^{PEG-FA}$ | - | - | 13 (± 2) | 13 (± 2) |
| R$_g^{PEG}$ | - | - | 6.4 (± 0.1) | 6.6 (± 0.1) |
| R$_g^{PEG-FA}$ | - | - | 8.4 (± 0.1) | 8.1 (± 0.1) |
| | Hydrogen bonds number | | | |
| NP-FA | 6.4 (± 0.9) | 2.1 (± 0.9) | 0 | 0 |
| NP-PEG | - | - | 0 | 0 |
| PEG-PEG | - | - | 1 (± 1) | 1 (± 1) |
| PEG-FA | - | - | 2 (± 1) | 3 (± 2) |
| FA-FA | 24 (± 6) | 16 (± 5) | 1 (± 1) | 4 (± 3) |
| FA-wat | 420 (± 12) | 422 (± 13) | 101 (± 7) | 198 (± 9) |
| PEG-wat | - | - | 348 (± 17) | 358 (± 13) |
| NP-wat | 24 (± 3) | 45 (± 4) | 76 (± 4) | 77 (± 4) |
| | Non-bonding interaction energies (kcal/mol) | | | |
| vdW NP-FA | -588 (± 10) | -348 (± 10) | -0.7 (± 0.7) | -0.2 (± 0.6) |
| ele NP-FA | -229 (± 11) | -99 (± 6) | 0.0 (± 0.3) | -0.3 (± 0.8) |
| vdW NP-PEG | - | - | -203 (± 11) | -178 (± 11) |
| ele NP-PEG | - | - | -84 (± 5) | -78 (± 4) |
| vdW PEG-PEG | - | - | -372 (± 23) | -364 (± 21) |
| ele PEG-PEG | - | - | 12 (± 14) | 9 (± 13) |
| vdW PEG-FA | - | - | -79 (± 17) | -140 (± 26) |
| ele PEG-FA | - | - | -40 (± 14) | -65 (± 22) |
| vdW FA-FA | -503 (± 23) | -421 (± 23) | -15 (± 7) | -75 (± 14) |
| ele FA-FA | -305 (± 33) | -245 (± 37) | -13 (± 13) | -63 (± 27) |
| vdW FA-wat | -684 (± 36) | -726 (± 37) | -177 (± 22) | -327 (± 36) |
| ele FA-wat | -4416 (± 127) | -4356 (± 124) | -1090 (± 57) | -2131 (± 94) |
| vdW PEG-wat | - | - | -1752 (± 54) | -1742 (± 56) |
| ele PEG-wat | - | - | -4200 (± 207) | -4287 (± 128) |
| vdW NP-wat | -113 (± 10) | -232 (± 16) | -150 (± 16) | -175 (± 15) |
| ele NP-wat | -577 (± 27) | -1038 (± 32) | -1155 (± 36) | -1172 (± 33) |
| vdW FA-solution | -682 (± 36) | -724 (± 37) | -176 (± 22) | -325 (± 35) |
| ele FA-solution | -4489 (± 124) | -4429 (± 120) | -1134 (± 52) | -2192 (± 94) |
| vdW PEG-solution | - | - | -1739 (± 58) | -1735 (± 57) |
| ele PEG-solution | - | - | -4703 (± 125) | -4588 (± 127) |
| vdW NP-solution | -113 (± 10) | -232 (± 16) | -150 (±16) | -175 (± 15) |
| ele NP-solution | -577 (± 27) | -1037 (± 32) | -1156 (± 36) | -1173 (± 33) |

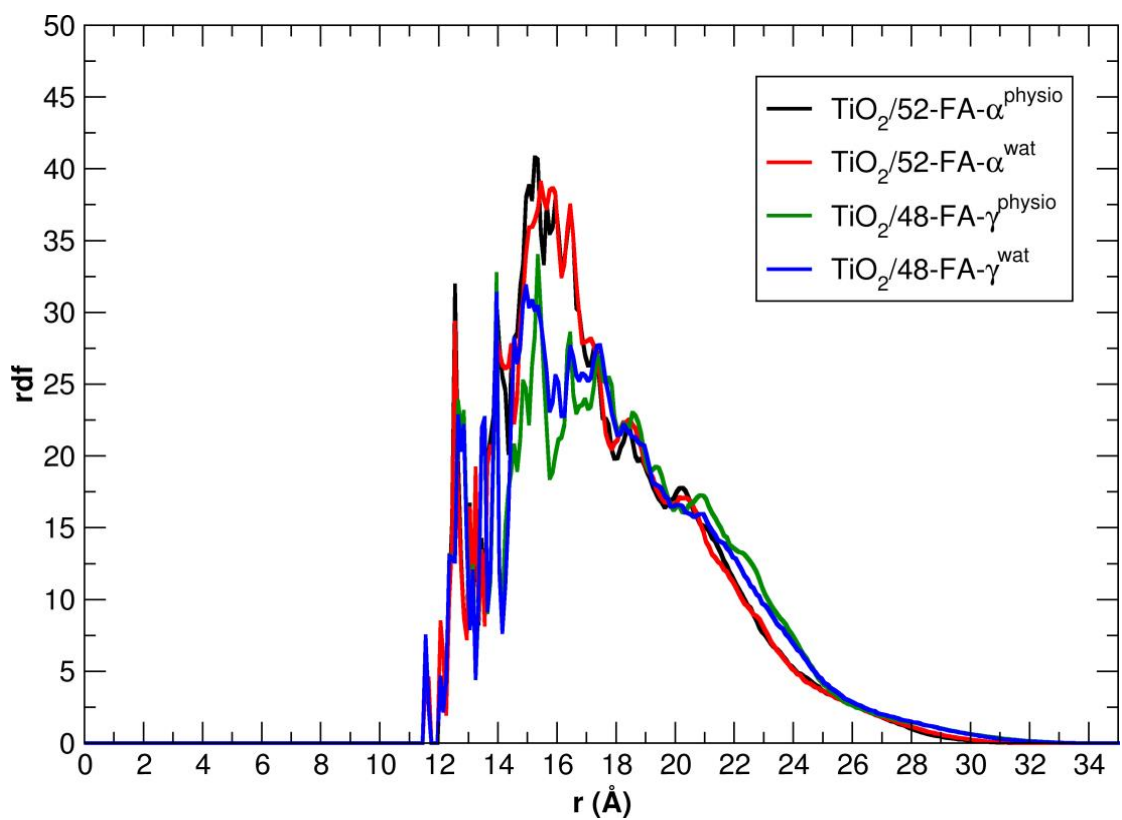

Fig. S13. Comparison of the radial distribution function (rdf) of the FA center of mass with respect to the six-fold coordinated Ti atom at the center of the NP, for the $TiO_2/52$-FA-$\alpha$ and $TiO_2/48$-FA-$\gamma$ systems, in pure water and under physiological conditions, averaged over the last 100 ns of MD production phase.

Table S8. Estimated NP+FA or NP+PEG+FA diffusion coefficients for the last 100 ns of MD production phase of the $TiO_2/52$-FA-$\alpha$, $TiO_2/48$-FA-$\gamma$, $TiO_2/PEG/10$-FA and $TiO_2/PEG/20$-FA systems in physiological environment.

| System | $D/10^{-10}$ (m$^2$/s) |
|---|---|
| $TiO_2/52$-FA-$\alpha$ | 1.7 ($\pm$ 0.1) |
| $TiO_2/48$-FA-$\gamma$ | 2.2 ($\pm$ 0.2) |
| $TiO_2/PEG/10$-FA | 1.9 ($\pm$ 0.2) |
| $TiO_2/PEG/20$-FA | 1.8 ($\pm$ 0.2) |